\documentclass[conference,compsoc]{IEEEtran}
\IEEEoverridecommandlockouts

\ifCLASSOPTIONcompsoc
  \usepackage[nocompress]{cite}
\else
  \usepackage{cite}
\fi

\usepackage{hyperref}
\hypersetup{pdftex,colorlinks=true,allcolors=blue}
\usepackage{tcolorbox}
\newtcolorbox{takeaway}{colframe=black,colback=gray!15,boxrule=1pt,arc=2pt,left=2pt,right=2pt,top=1pt,bottom=1pt, before skip=1em, after skip=0em}
\usepackage{url}
\usepackage[normalem]{ulem}

\AtBeginDocument{%
	\providecommand\BibTeX{{%
			\normalfont B\kern-0.5em{\scshape i\kern-0.25em b}\kern-0.8em\TeX}}}

\usepackage{setspace}
\usepackage{longtable}
\usepackage[figuresright]{rotating}
\usepackage{epsfig,endnotes,color}
\usepackage{graphicx}
\usepackage{amsmath}
\usepackage{caption}
\usepackage{graphicx}
\usepackage{url}
\usepackage{xcolor}
\usepackage{transparent}
\usepackage{float}
\usepackage{lipsum}
\usepackage{listings}
\definecolor{codegreen}{rgb}{0,0.6,0}
\definecolor{codegray}{rgb}{0.5,0.5,0.5}
\definecolor{codepurple}{rgb}{0.58,0,0.82}
\definecolor{backcolour}{rgb}{0.95,0.95,0.95}
\definecolor{verylightgray}{rgb}{.97,.97,.97}
\definecolor{custompink}{RGB}{216,100,139}
\definecolor{lightblue}{rgb}{0.4, 0.6, 0.9}
\definecolor{chocolate}{rgb}{0.36, 0.2, 0.09}
\definecolor{darkbrown}{rgb}{0.3, 0.15, 0.05}
\definecolor{saddlebrown}{HTML}{8B4513}
\definecolor{navyblue}{rgb}{0.0,0.0,0.5}

\lstdefinelanguage{Solidity}{
	keywords=[1]{anonymous, assembly, assert, balance, break, call, callcode, case, catch, class, constant, continue, constructor, contract, debugger, default, delegatecall, delete, do, else, emit, event, experimental, export, external, false, finally, for, function, gas, if, implements, import, in, indexed, instanceof, interface, internal, is, length, library, log0, log1, log2, log3, log4, memory, modifier, new, payable, pragma, private, protected, public, pure, push, require, return, returns, revert, selfdestruct, send, solidity, storage, struct, suicide, super, switch, then, this, throw, transfer, true, try, typeof, using, value, view, while, with, addmod, ecrecover, keccak256, mulmod, ripemd160, sha256, sha3}, 
	keywordstyle=[1]\color{blue}\bfseries,
	keywords=[2]{address, bool, byte, bytes, bytes1, bytes2, bytes3, bytes4, bytes5, bytes6, bytes7, bytes8, bytes9, bytes10, bytes11, bytes12, bytes13, bytes14, bytes15, bytes16, bytes17, bytes18, bytes19, bytes20, bytes21, bytes22, bytes23, bytes24, bytes25, bytes26, bytes27, bytes28, bytes29, bytes30, bytes31, bytes32, enum, int, int8, int16, int24, int32, int40, int48, int56, int64, int72, int80, int88, int96, int104, int112, int120, int128, int136, int144, int152, int160, int168, int176, int184, int192, int200, int208, int216, int224, int232, int240, int248, int256, mapping, string, uint, uint8, uint16, uint24, uint32, uint40, uint48, uint56, uint64, uint72, uint80, uint88, uint96, uint104, uint112, uint120, uint128, uint136, uint144, uint152, uint160, uint168, uint176, uint184, uint192, uint200, uint208, uint216, uint224, uint232, uint240, uint248, uint256, var, void, ether, finney, szabo, wei, days, hours, minutes, seconds, weeks, years},	
	keywordstyle=[2]\color{teal}\bfseries,
	keywords=[3]{block, blockhash, coinbase, difficulty, gaslimit, number, timestamp, msg, data, gas, sender, sig, value, now, tx, gasprice, origin},	
	keywordstyle=[3]\color{violet}\bfseries,
	identifierstyle=\color{black},
	sensitive=false,
	comment=[l]{//},
	morecomment=[s]{/*}{*/},
	commentstyle=\color{codegreen}\ttfamily,
	stringstyle=\color{red}\ttfamily,
	morestring=[b]',
	morestring=[b]"
}

\definecolor{hl}{RGB}{255,248,220}

\lstdefinestyle{soltight}{
  language=Solidity,
  frame=none,
  linewidth=.48\textwidth,
  xleftmargin=.03\textwidth,
  aboveskip=0pt,    
  belowskip=0pt     
}

\usepackage{svg}
\usepackage{CJKutf8}
\usepackage{subfigure}
\usepackage{lscape, latexsym, amssymb, algorithmic, multirow}
\usepackage[linesnumbered, vlined, ruled]{algorithm2e}
\usepackage{multirow}
\usepackage{mathtools, bbm, color}
\usepackage{booktabs}
\usepackage{amsthm,mathrsfs,amsfonts,dsfont}
\usepackage{enumerate}
\usepackage{hhline}

\usepackage{enumitem}
\usepackage{tabu}
\usepackage{colortbl}
\usepackage[graphicx]{realboxes}

\usepackage{footnote}
\usepackage{color}
\usepackage{colortbl}
\usepackage{pifont}
\usepackage{bbding}
\usepackage{threeparttable}
\usepackage{rotating}

\usepackage{marvosym}

\def\BibTeX{{\rm B\kern-.05em{\sc i\kern-.025em b}\kern-.08em
    T\kern-.1667em\lower.7ex\hbox{E}\kern-.125emX}}

\newcommand{\solcode}[1]{\textcolor{saddlebrown}{#1}}

\newcommand{\system}{{\sc SCRutineer}\xspace}
\newcommand{\aguke}{{\sc UkeAgent}\xspace}
\newcommand{\agluc}{{\sc LuvAgent}\xspace}

\newfont{\mycrnotice}{ptmr8t at 7pt}

\IEEEaftertitletext{\vspace{1.5\baselineskip}}

\begin{document}

\title{\system: Detecting Logic-Level Usage Violations of Reusable Components in Smart Contracts}

\author{
\IEEEauthorblockN { 
Xingshuang Lin\IEEEauthorrefmark{1},
Binbin Zhao\IEEEauthorrefmark{1},
Jinwen Wang\IEEEauthorrefmark{1},
Qinge Xie\IEEEauthorrefmark{2},
Xibin Zhao\IEEEauthorrefmark{3},
Shouling Ji\IEEEauthorrefmark{1}
}
\IEEEauthorblockA {
  \IEEEauthorrefmark{1}Zhejiang University,
  \IEEEauthorrefmark{2}Georgia Institute of Technology, 
  \IEEEauthorrefmark{3}Tsinghua University
}
\IEEEauthorblockA {
  E-mails: 
  cs.xslin@zju.edu.cn, 
  binbinz@zju.edu.cn,
  jinwen.22@intl.zju.edu.cn,
  qxie47@gatech.edu, \\
  zxb@tsinghua.edu.cn,
  sji@zju.edu.cn
}
}

\maketitle


\begin{abstract} 

Smart Contract Reusable Components (SCRs) play a vital role in accelerating the development of business-specific contracts by promoting modularity and code reuse. 
However, the risks associated with SCR usage violations have become a growing concern. One particular type of SCR usage violation, known as a logic-level usage violation, is becoming especially harmful. This violation occurs when the SCR adheres to its specified usage rules but fails to align with the specific business logic of the current context, leading to significant vulnerabilities. Detecting such violations necessitates a deep semantic understanding of the contract’s business logic, including the ability to extract implicit usage patterns and analyze fine-grained logical behaviors.

To address these challenges, we propose \system, the first automated and practical system for detecting logic-level usage violations of SCRs. 
First, we design a composite feature extraction approach that produces three complementary feature representations, supporting subsequent analysis. 
We then introduce a Large Language Model (LLM)-powered knowledge construction framework, which leverages comprehension-oriented prompts and domain-specific tools to extract logic-level usage and build the SCR knowledge base. 
Next, we develop a Retrieval-Augmented Generation (RAG)-driven inspector, which combines a rapid retrieval strategy with both comprehensive and targeted analysis to identify potentially insecure logic-level usages. 
Finally, we implement a logic-level usage violation analysis engine that integrates a similarity-based checker and a snapshot-based inference conflict checker to enable accurate and robust detection. 
We evaluate \system from multiple perspectives on $3$ ground-truth datasets. The results show that \system achieves a precision of \textbf{$80.77\%$}, a recall of \textbf{$82.35\%$}, and an F1-score of \textbf{$81.55\%$} in detecting logic-level usage violations of SCRs. Moreover, we conduct a real-world analysis on $1,181$ on-chain contracts, successfully identifying \textbf{$13$} previously undocumented logic-level usage violations of SCRs, which can result in vulnerabilities. As of now, $9$ of these vulnerabilities have been assigned CVE IDs.
Notably, the zero-day vulnerabilities identified in one of the most prominent DeFi platforms, with a market capitalization exceeding $\$1$ billion and a 24-hour trading volume surpassing $\$150$ million, have been confirmed by their security teams.

\end{abstract}


\section{Introduction}
\label{Section:intro}

Smart contracts are self-executing digital agreements deployed on blockchain platforms, with predefined rules that facilitate trustless and automated transactions without intermediaries. 
Smart contracts are predominantly implemented in Solidity, the most widely used programming language for Ethereum and other EVM-compatible blockchains. They serve as the foundational infrastructure for decentralized applications across various blockchain ecosystems. In modern smart contract development, Smart Contract Reusable Components (SCRs) significantly streamline the construction of business-specific contracts. SCRs are modular code units, such as interfaces, abstract contracts, libraries, and base contracts, that encapsulate general-purpose logic and can be imported, inherited, or cloned by business contracts~\cite{fse23/sun}.

A recent study~\cite{icse/HuangCJZ24} reveals that approximately 70\% of smart contracts incorporate SCRs, highlighting their widespread adoption in practice. 
However, the risks associated with SCR usage violations have become a growing concern. One particular type of SCR usage violation, known as logic-level usage violation, is becoming increasingly detrimental.
This violation exhibits distinct characteristics compared to those in traditional software libraries. Specifically, usage violations in traditional libraries typically stem from breaches of explicitly defined syntactic rules, referred to as syntax-level usage violations. In contrast, most usage violations involving SCRs often relate to logic-level usage violations, which arise from implicit usage constraints that are not explicitly documented in the SCR specifications. These implicit constraints are closely tied to the underlying business logic of smart contracts and must be analyzed across various scenarios. Logic-level usage violations of SCRs can lead to significant vulnerabilities, potentially resulting in substantial financial losses.

Currently, many works have focused on smart contract security, which can be categorized into two main directions based on their analytical perspectives. The first line of work focuses on the detection of contract-level vulnerabilities. These approaches have demonstrated strong effectiveness in identifying common vulnerability types such as reentrancy, improper permission validation, and others~\cite{ndss/propertygpt2025, shou2023ityfuzz, sp/smartinv2024}.
The second direction centers on the security analysis of contract reuse. Prior efforts in this area have made progress in detecting the misuse of reusable components~\cite{icse/HuangCJZ24}, identifying vulnerabilities in component variants~\cite{uss/zepscope2025}, and analyzing security risks in cross-chain component reuse~\cite{issta/equivguard2025}.
Despite these advances, logic-level usage violations of SCRs remain largely unaddressed. For instance, in 2024, the ATM token on the BNB Chain suffered an exploit due to logic-level usage violations of SCRs, resulting in a financial loss of approximately \$180,000~\cite{ATM/BlockSec}.
The core limitation of existing detection tools lies in their focus on either syntax-level usage violations or general contract vulnerabilities, without the ability to detect logic-level usage violations. In practice, identifying such violations requires a deeper semantic understanding of the contract source code and the ability to extract implicit usage patterns of SCRs, capabilities that are beyond the scope of current methodologies.
This gap highlights the pressing need for a practical and automated system capable of detecting logic-level usage violations of SCRs, thereby addressing an important yet underexplored aspect of smart contract security.

\subsection{Challenges}
To detect logic-level usage violations of SCRs automatically, we have the following key challenges.

\textbf{Challenge I: Inference of Relevant Logic-Level Usage Knowledge of SCRs.}
Detecting logic-level usage violations of SCRs requires a comprehensive understanding of their intended logic-level usage. The first challenge lies in the automated inference of relevant logic-level usage knowledge, which is a foundational step for enabling accurate violation analysis. This task is particularly difficult because most SCRs are distributed as source code with limited or no accompanying documentation. Even for well-documented components, the implicit nature of logic-level usage constraints poses a substantial challenge for automated inference.

\textbf{Challenge II: Identification of Potentially Insecure Logic-Level Usages of SCRs.}
The logic-level usage knowledge inferred from SCRs is often abstract and cannot be directly applied to analysis scenarios. The second challenge, therefore, is to bridge the gap between inferred usage knowledge and effective analysis by designing an effective inspector. 
This process entails interpreting abstract logic-level usage knowledge, translating it into actionable criteria, and analyzing contract scenarios to identify potentially insecure logic-level usages of SCRs.
The task is nontrivial, as it involves multi-step reasoning over the contextual constraints of SCRs and the business logic of smart contracts. Furthermore, the inspector must analyze multi-dimensional contract features, making both the reasoning strategy and the construction of an effective reasoning pipeline significantly more complex.

\textbf{Challenge III: Accurate Detection of Logic-Level Usage Violation of SCRs.}
Even after identifying potentially insecure logic-level usages of SCRs, the analysis may still suffer from false positives due to the coarse granularity of multi-dimensional reasoning. As such, effective detection of logic-level usage violations of SCRs requires support from fine-grained, context-aware analysis mechanisms. The third challenge, therefore, is to design a robust and practical detection method capable of validating suspected violations with higher precision. This necessitates the integration of fine-grained structural insights, logic-level usage features, and mitigation of false inferences propagated from earlier phases of reasoning.

\subsection{Methodology} 

In this paper, we aim to address these challenges to detect logic-level usage violations of SCRs. To this end, we propose \system, an automated and practical system to conduct SCR logic-level usage violation detection in smart contracts. Our design philosophy is as follows.

\textbf{First}, to support automated SCR usage analysis, we design a composite feature extraction approach that captures critical features of SCR usages. This approach is built upon a tri-component compression strategy that produces three complementary representations: a composite signature, a logical sequence, and a latent structural embedding. These representations are dynamically adaptable and can be selectively applied based on the requirements of downstream analysis.
\textbf{Next}, to address \textbf{Challenge I}, we propose a Large Language Model (LLM)-powered framework for constructing the SCR knowledge base. This framework operates in two stages. In the first stage, we implement a targeted crawler that collects SCRs from diverse sources, prioritizing those with high usage frequency and a history of associated violation incidents. In the second stage, we deploy an SCR analysis agent, \aguke, which is equipped with a task-planning middleware, comprehension-oriented prompts, and domain-specific tools. \aguke facilitates accurate and robust extraction of logic-level usage knowledge of SCRs.
\textbf{Then}, to address \textbf{Challenge II}, we develop a Retrieval-Augmented Generation (RAG)-driven inspector, \agluc, which integrates two core strategies: a retrieval strategy that efficiently locates relevant reference data from the knowledge base, and an augmented generation strategy that performs both comprehensive and targeted analyses. Together, these strategies enable the detection of potentially insecure logic-level usages.
\textbf{Finally}, to address \textbf{Challenge III}, we implement a logic-level usage violation analysis engine that integrates two complementary techniques: a similarity-based checker and a snapshot-based inference conflict checker. This multi-layered design enables effective and reliable detection of logic-level usage violations of SCRs.
By integrating the above components into a unified system, we achieve high precision, recall, and F1-score in detecting logic-level usage violations of SCRs.

\subsection{Contribution}

We summarize our main contributions as follows:

$\bullet$ To the best of our knowledge, we propose \system, the first automated and practical system for detecting logic-level usage violations of SCRs, bridging a critical gap in existing smart contract security research. Furthermore, we construct the first ground-truth dataset of verified logic-level usage violation cases, encompassing \textbf{$382$} smart contracts collected from the real-world attack incidents. To support future research and facilitate reproducibility, we will release the full implementation of \system at \url{https://anonymous.4open.science/r/SCRUTINEER-97CF} upon paper acceptance.

$\bullet$ We design an SCR analysis agent, \aguke, which is equipped with a task-planning middleware, comprehension-oriented prompts, and domain-specific tools to automatically construct a reliable SCR usage knowledge base. Our RAG-driven inspector, \agluc, integrates retrieved knowledge with both comprehensive and targeted analyses to identify potentially insecure logic-level usages. Moreover, the detection engine employs two tailored analysis strategies that ensure effective and reliable detection.

$\bullet$ We implement and evaluate \system on ground-truth datasets. The results show that \system achieves a precision of \textbf{$80.77\%$}, a recall of \textbf{$82.35\%$}, and an F1-score of \textbf{$81.55\%$} in detecting logic-level usage violations of SCRs, significantly outperforming existing approaches.

$\bullet$ Our real-world analysis of \textbf{$1,181$} on-chain contracts demonstrates the effectiveness of \system in detecting previously overlooked logic-level usage violations of SCRs.
\system identifies $13$ zero-day vulnerabilities, $9$ of which have been assigned CVE IDs. Notably, the zero-day vulnerabilities identified in one of the most prominent DeFi platforms, with a market capitalization exceeding $\$1$ billion and a 24-hour trading volume surpassing $\$150$ million, have been confirmed by their security team.

\section{Background}

In this section, we provide a brief introduction to retrieval-augmented generation, usage violations, and a motivating example.

\subsection{Retrieval-Augmented Generation}

To analyze logic-level SCR usages, \system often requires context about how an SCR is intended to be used and aligned with its logic constraints. Retrieval-Augmented Generation (RAG) enables \system to obtain such information by retrieving relevant SCR instances from the knowledge base and grounding its inference.

RAG is a hybrid framework that supplements LLMs with external knowledge retrieved from a vector database, thereby enhancing the in-context reasoning capability of LLMs~\cite{DBLP:conf/emnlp/JiangXGSLDYCN23}~\cite{DBLP:conf/nips/LewisPPPKGKLYR020}. Several mainstream agent development frameworks, such as LangChain~\cite{langchain}, RAGflow~\cite{ragflow}, and LlamaIndex~\cite{llama_index}, have integrated support for building RAG systems.
A typical RAG system follows a two-stage pipeline. In the first stage, the input query is encoded into a dense vector representation and used to retrieve semantically relevant documents from a pre-constructed vector database. In the second stage, the retrieved content is incorporated into the prompt context of the LLM, enabling it to generate more accurate and informed responses.

RAG systems have demonstrated substantial progress in a wide range of domains, including code generation~\cite{DBLP:conf/kbse/GaoXWGSW024}, automated program repair~\cite{DBLP:conf/icse/OuyangZ0M25}, and vulnerability detection~\cite{DBLP:conf/aaai/Li0LW25}. By incorporating external domain-specific knowledge into the reasoning process~\cite{DBLP:conf/aaai/ChowdhuryZW22}, RAG enhances LLM performance along three key dimensions.
First, RAG enables the generation of responses grounded in up-to-date domain knowledge, significantly improving both accuracy and robustness~\cite{DBLP:conf/emnlp/YuWCZ24}.
Second, hallucinations, which are confident yet incorrect outputs, are a well-known limitation of LLMs. RAG mitigates this issue by grounding responses in verifiable external knowledge retrieved from curated sources~\cite{DBLP:conf/aaai/0011LH024}.
Third, RAG enhances resource efficiency. Instead of retraining or fine-tuning large models as datasets evolve, developers can simply expand the underlying vector store and rebuild the retrieval index. This lightweight update mechanism reduces computational overhead and lowers associated operational costs~\cite{DBLP:conf/acl/XuPYMSCZ24}~\cite{DBLP:conf/acl/WangTOXS24}.

\subsection{Usage Violations of SCRs}

As SCRs provide standardized building blocks for smart contract development, their correct usage is essential for preventing risky contract behaviors. 
In practice, securely using SCRs requires attention to two types of usage violations: syntax-level usage violations and logic-level usage violations.

\begin{itemize}[nosep, leftmargin=*]
    \item[$\bullet$] \textbf{Syntax-level Usage Violations.}
In smart contracts, a syntax-level usage violation refers to a violation arising from the misinterpretation or misunderstanding of the syntactic rules in SCR specifications.
Specifically, such violations may involve passing incorrect out-of-scope arguments to SCRs~\cite{bg/param}, misusing their return values~\cite{bg/returntype}, invoking inappropriate SCRs or overestimating their functional capabilities~\cite{icse/HuangCJZ24}. 

  \item[$\bullet$] \textbf{Logic-Level Usage Violations.}
In smart contracts, a logic-level usage violation refers to a violation arising from the neglect of implicit usage constraints of SCRs that are not explicitly documented in their specifications. These implicit usage constraints are often context-dependent, closely tied to the business logic of smart contracts, and require dynamic reasoning across different scenarios.
In this paper, we primarily focus on addressing logic-level usage violations of SCRs, as mentioned in Section~\ref{Section:intro}.
\end{itemize}

\subsection{Motivating Example}
\label{motivation}

\begin{figure}[h]
\begin{lstlisting}[
      style=soltight,
      firstline=1, lastline=10,
      firstnumber=1
    ]
contract SimpleContract{
  IUniswapV2Router02 public _swapRouter;
  TokenDistributor public _tokenDistributor;
  address public currency;
  function swapTokensForCurrency(uint256 tokenAmount) private {
    address[] memory path = new address[](2);
    path[0] = address(this);
    path[1] = currency;
    _approve(address(this), address(_swapRouter), tokenAmount);
    try
\end{lstlisting}
    {\begingroup
    \lstset{backgroundcolor=\color{hl}, firstnumber=last}
\begin{lstlisting}[
      style=soltight,
      firstline=11, lastline=11
    ]
      _swapRouter.swap(tokenAmount,0,path,address(_tokenDistributor),block.timestamp)
\end{lstlisting}
    \endgroup}
\begin{lstlisting}[
      style=soltight,
      firstline=12,
      firstnumber=last
    ]
    {} catch { emit Failed_swap(1); }
    uint256 currencyBal = IERC20(currency).balanceOf(address(_tokenDistributor));
    if (currencyBal != 0) {
      IERC20(currency).transferFrom(address(_tokenDistributor),address(this),currencyBal); } } }
\end{lstlisting}
    \caption{The violation code snippet in the DeFiHackLabs case (line 11).}
    \label{Figure: Motivating Example}
\end{figure}

We provide an SCR logic-level usage violation discovered on DeFiHackLabs~\cite{defihacklabs}, the case of misaligned parameter logic resulting from an implicit usage constraint, as a motivating example, as shown in Figure~\ref{Figure: Motivating Example}.

In the \solcode{SimpleContract}, the function \solcode{swapTokensForCurrency} swaps tokens by calling the SCR \solcode{\_swapRouter.swap(...)} derived from \textit{Uniswap}~\cite{uniswap}. The logic-level usage violation arises from setting the second parameter of \solcode{\_swapRouter.swap(...)} to a fixed value of zero, which denotes the minimum acceptable output amount. Despite being syntactically correct, the second parameter must be logically coupled with the first parameter, \solcode{tokenAmount} (the input amount). In real token-swap business logic, the minimum output must scale with the input to protect against slippage and price manipulation. By breaking this implicit constraint, the contract unintentionally allows an attacker to execute a sandwich attack, manipulating prices so the victim receives significantly fewer tokens.

While prior studies~\cite{icse/HuangCJZ24} have investigated the security of SCRs, they fail to detect such issues. The core limitation lies in their inability to capture and reason about implicit logic-level usage patterns, as well as their lack of understanding of contract-specific business semantics.

To reveal such implicit violations, \system employs a unified analysis pipeline that integrates contract feature extraction, knowledge base construction, retrieval-augmented inspector, and a violation detection engine. Through these modules, \system captures hidden logic constraints like the one violated in this example and reliably detects the logic-level usage violation of SCRs.

\section{\system Design}

\begin{figure*}[h]
\centering
\includegraphics[width=0.96\textwidth]{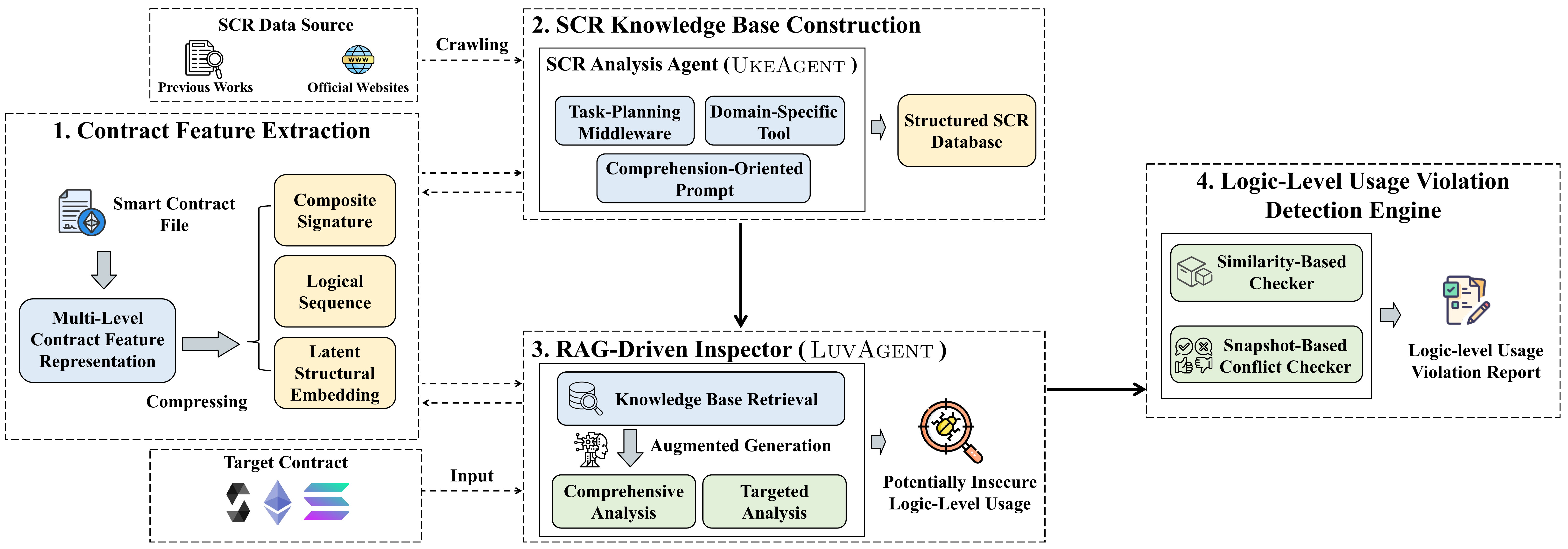}
\caption{Framework of \system.}
\label{Figure: system framework}
\end{figure*}

In this section, we present the detailed design of \system, which aims to automatically detect logic-level usage violations of SCRs within smart contracts. As shown in Figure~\ref{Figure: system framework}, \system comprises the following four key modules: 
\begin{itemize}[nosep, leftmargin=*]
    \item[$\bullet$] \textbf{Contract Feature Extraction}:  \system first utilizes the contract feature extraction module to analyzes the input smart contract to construct multi-level feature representations that serve as the foundation for subsequent analysis. These representations include a composite signature, a logical sequence, and a latent structural embedding, each capturing the contract’s underlying logic and behavioral structure at different levels of abstraction.
    \item[$\bullet$] \textbf{SCR Knowledge Base Construction}: Next, the SCR knowledge base construction module is used to collect SCR-related content from diverse sources and constructs a structured knowledge base. This process is guided by an SCR analysis agent, \aguke, equipped with a task-planning middleware, comprehension-oriented prompts, and domain-specific tools, ensuring accurate parsing, normalization, and integration of SCR information.
    \item[$\bullet$] \textbf{RAG-Driven Inspector}: the RAG-driven inspector module, \agluc , enhances the LLM’s reasoning capabilities by querying the SCR knowledge base to retrieve relevant reference data. It employs an RAG strategy, integrated with both comprehensive and targeted analyses, to identify potential insecure logic-level usages of SCRs.
    \item[$\bullet$] \textbf{Logic-Level Usage Violation Detection Engine}: the logic-level usage violation detection engine conducts a fine-grained analysis of the potential cases. It employs two complementary techniques: a similarity-based checker and a snapshot-based inference conflict checker. These techniques work in concert to validate potential logic-level usage violations of SCRs within smart contracts. 
\end{itemize}

\subsection{Contract Feature Extraction}
\label{Smart Contract Feature Extraction}

The purpose of smart contract feature extraction is to identify both explicit and implicit features from the contract and abstract them into a structured representation that supports subsequent analysis. Detecting logic-level usage violations of SCRs necessitates leveraging multi-dimensional program features, as such issues cannot be accurately identified from a single perspective alone. Nevertheless, it is challenging to construct a precise and informative feature representation. The main problem lies in balancing conciseness with expressiveness, ensuring the representation is compact while still capturing the unique structural and semantic properties of the contract. To address the above challenges, we propose a composite analysis approach for contract feature extraction, incorporating the following designs.

\noindent\textbf{Multi-Level Contract Feature Representation}.
Performing in-depth analysis directly at the source-code level often yields noisy or incomplete information, as smart contract source codes lack the structural annotations required for reliable program analysis. To obtain a more comprehensive representation, we construct a composite feature graph for each smart contract. We first generate the Abstract Syntax Tree (AST) of the smart contract by compiling its source code. To ensure compatibility with different Solidity versions, we design and implement an adaptive contract compiler capable of automatically identifying the Solidity version specified in the source code and configuring the corresponding compiler, thereby enabling accurate and efficient AST construction. Based on the generated AST, we extract three fundamental program graphs, including the Program Call Graph (PCG), the Data Flow Graph (DFG), and the Control Flow Graph (CFG). We then integrate these graphs into a unified multi-level composite feature graph that holistically captures the contract’s structural and behavioral characteristics.

\noindent\textbf{SCR Feature Extraction}.
Given the composite graph representation constructed for each contract, it becomes feasible to extract and analyze the SCR usage within smart contracts. However, directly utilizing the complete composite graph introduces substantial computational overhead, which can degrade system performance. Therefore, an effective graph compression strategy is critical. 
In designing this strategy, the compressed representation must strike a balance between efficiency and informativeness. It should be compact enough to enable efficient matching and retrieval within the SCR knowledge base, while still preserving the essential logic-level usage features and internal structural features. Therefore, we propose a tri-component compression strategy that extracts the composite signature, the logical sequence, and the latent structural embedding. These representations can be used individually or in combination to support the subsequent logic-level usage violation analysis with both flexibility and efficiency.

\begin{figure}
\centering
\includegraphics[width=0.48\textwidth]{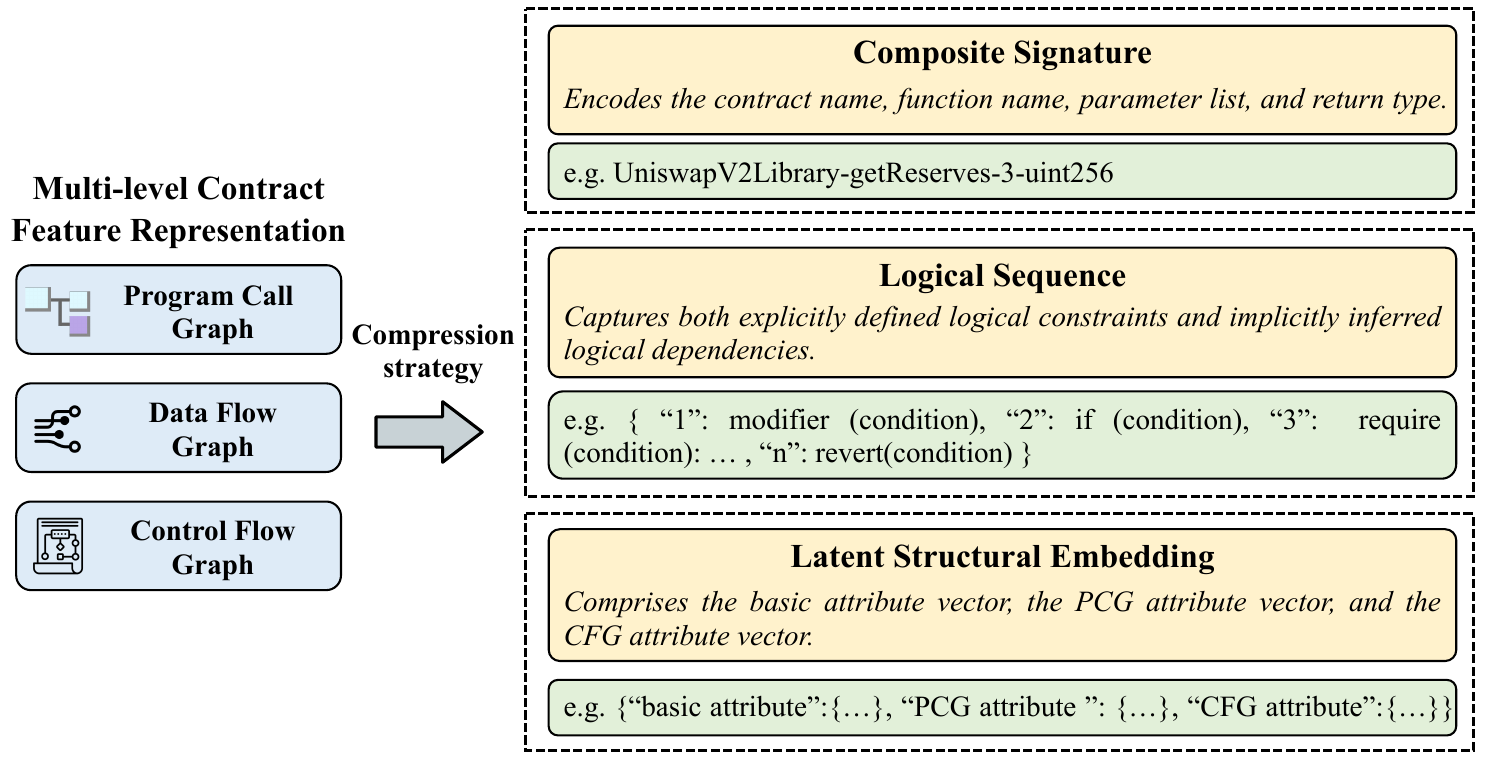}
\caption{Feature compression process.}
\label{Figure: Feature compression process}
\end{figure}

\textbf{1) Composite Signature for Efficient Matching}.
Since the composite signature is primarily utilized in scenarios requiring rapid matching, it is intentionally designed to retain only essential information related to SCR usage. Specifically, it encapsulates the contract name, function name, parameter list, and return types. 
For instance, as shown in Figure~\ref{Figure: Feature compression process}, we encode the function \texttt{getReserves(address factory, address tokenA, address tokenB)} in the contract \texttt{UniswapV2Library} with a return type \texttt{uint256} to \texttt{UniswapV2Library-getReserves-3-uint256}.
This compact representation enables fast retrieval and facilitates efficient message passing between system modules.

\textbf{2) Logical Sequence for Capturing Implicit Constraints}.
The logic-level usage feature associated with SCR usage is often entangled with various structural and semantic elements of the smart contract.
In our approach, we define the logical sequence to represent the logic-level usage feature. The logical sequence consists of an SCR usage along with its surrounding logical constraints. The algorithm traverses each function node related to SCR usage, examining its structural and semantic context, including both component calls, inheritance, and overriding. For SCR calls, we examine control-flow constructs such as \texttt{modifier}, \texttt{if-else}, \texttt{require}, \texttt{revert}, and other relevant nodes to infer the explicit logical constraints that govern the invocation of the SCR.
For SCR inheritance and overriding, we analyze both inherited contracts and overridden functions to identify behavioral differences. These differences are then used to infer implicit logical constraints that affect SCR usage in the context of polymorphic behavior. For instance, as shown in Figure \ref{Figure: Feature compression process}, the logical sequence is represented in the following format: \{``1'', ``\texttt{modifier(condition)}'', ``2'': ``\texttt{if(condition)}'', ``3'': ``\texttt{require(condition)}'', ... , ``n'': ``\texttt{revert(condition)}''\}.

\textbf{3) Latent Structural Embedding for Internal Analysis}. The internal structural features of an SCR are also crucial for understanding how its behavior emerges from underlying structural patterns and analyzing the usage of SCRs.
In our approach, we categorize SCR structural attributes into three categories: basic attributes, PCG attributes, and CFG attributes. The basic attributes capture essential contract-level information, including the number of nodes, the number of parameters, and the return type. The PCG attributes describe inter-component relationships, such as the number of internal function calls and external calls. The CFG attributes characterize the fine-grained internal control flow structure, including metrics such as the number of entry-point nodes, variable nodes, expression nodes, conditional nodes, loop nodes, and return nodes. For instance, as shown in Figure \ref{Figure: Feature compression process}, the latent structural embedding of \texttt{getReserves(...)} is represented in the following format: \{``\texttt{basic attribute}'': ... , ``\texttt{PCG attribute}'': ... , ``\texttt{CFG attribute}'': ... \}.
This structured attribute-based compression enables efficient representation and comparison of SCR internal logic.

\subsection{SCR Knowledge Base Construction}
\label{SRC Knowledge Base Construction}

The purpose of SCR knowledge base construction is to collect SCR code from diverse sources and build a structured reference database to support subsequent violation analysis. Detecting logic-level usage violations of SCRs requires a comprehensive understanding of their intended logic-level usage specifications. 
Nevertheless, it is challenging to design an automated approach to infer the concealed logic-level usage knowledge of SCRs. 
The main problem lies in two aspects.
First, the sheer number and diversity of SCRs used across smart contracts make it difficult to devise an effective collection strategy. Second, most SCRs are distributed as source code with limited or no accompanying documentation. Even when documentation exists, it tends to emphasize syntax-level usage correctness rather than deeper logic-level usage specifications.
To address these challenges, we propose an LLM-powered construction framework that integrates a tailored crawling mechanism with an SCR analysis agent. The framework consists of two core stages: SCR Collection and SCR Analysis.

\noindent\textbf{SCR Collection.} 
In this stage, we collect SCR source codes and documents to construct a raw dataset. Nevertheless, it is challenging to download SCRs directly since there is no unified access channel available. To solve the problem of dispersed data resources, we implement a crawler equipped with two strategies to collect SCRs. Specifically, we determine the first data source based on usage frequency. Previous research has analyzed the usage of libraries in real-world Ethereum smart contracts~\cite{icse/HuangCJZ24}. They find $3,864$ types of libraries used in $156,761$ smart contracts, and rank all libraries based on their usage frequency. We take the Top $40$ libraries among them as one of our data sources\cite{icse/HuangCJZ24}. Additionally, we determine the second data source based on historical logic-level usage violation incidents.

\begin{table*}
\centering
\caption{Description for each plugin, along with the associated plugin type.}
\label{Table: Description for each plugin.}
\resizebox{0.96\textwidth}{!}{
\begin{tabular}{clllc}
\toprule[1.5pt]
\textbf{Plugin Type}                       & \textbf{Plugin Name}    & \textbf{Description}  & \textbf{Parameter} & \textbf{Return Type} \\ \midrule[1pt]

\multirow{8}{*}{\begin{tabular}[c]{@{}c@{}} Tool-invoking\\Plugin \end{tabular}}  
& \texttt{complie\_solidity\_contract}  & Complie the specified smart contract   &  \begin{tabular}[c]{@{}l@{}} The solidity file path   \\    \end{tabular}  & Class \\ \cmidrule{2-5}
& \texttt{read\_specified\_range}      & Read a range of lines in a solidity file   & \begin{tabular}[c]{@{}l@{}} The array containing line numbers\end{tabular}  & List \\ \cmidrule{2-5}
& \texttt{get\_all\_contracts}        & Get all contract instances in a solidity file    & \begin{tabular}[c]{@{}l@{}}  The intermediate representation of solidity \end{tabular}   &  List  \\  \cmidrule{2-5}
& \texttt{judge\_interface}  & Assess whether a contract has been implemented & \begin{tabular}[c]{@{}l@{}}  The instance of contract class \\    \end{tabular} & Bool \\ \cmidrule{2-5} 
& \texttt{get\_all\_functions\_by\_contract}   & Get all function instances in the specified contract  &  \begin{tabular}[c]{@{}l@{}}   The instance of contract class  \end{tabular}  & List   \\ \cmidrule{2-5}
& \texttt{extract\_calls\_by\_function}  & Get all calls in the specified function & \begin{tabular}[c]{@{}l@{}}  The instance of function class  \\    \end{tabular} & List \\ \cmidrule{2-5} 
& \texttt{extract\_CFG\_by\_function}   &  Get the control flow graph in the specified function &  \begin{tabular}[c]{@{}l@{}}  The instance of function class  \\    \end{tabular} &  JSON\\ \midrule[1pt]
\multirow{6}{*}{\begin{tabular}[c]{@{}c@{}} Response-processing \\ Plugin \end{tabular}}     
& \texttt{json\_data\_parsing}   & \begin{tabular}[c]{@{}l@{}} Parse JSON format data generated by the tools \\ using the specified strategy \end{tabular}  & \begin{tabular}[c]{@{}l@{}}  1) JSON format data \\  2) Strategy configuration file  \end{tabular}   & Map   \\ \cmidrule{2-5} 
& \texttt{list\_data\_parsing}   & \begin{tabular}[c]{@{}l@{}} Parse List format data generated by the tools \\ using the specified strategy  \end{tabular}  & \begin{tabular}[c]{@{}l@{}}  1) List format data  \\  2) Strategy configuration file  \end{tabular}   & Map     \\ \cmidrule{2-5} 
& \texttt{exception\_parsing}     & \begin{tabular}[c]{@{}l@{}} Parse the exceptions generated by the tools and \\ generate the prompt using prompt template  \end{tabular} & \begin{tabular}[c]{@{}l@{}}  1) Exception data \\  2) Prompt template ID  \end{tabular}  & String             \\ \bottomrule[1.5pt]
\end{tabular}
}
\end{table*}

\noindent\textbf{SCR Analysis.}
In this stage, we aim to extract precise logic-level usage knowledge of SCRs from the corresponding source code. The knowledge is subsequently converted into structured representations and utilized by both the RAG-driven inspector and the logic-level usage violation detection engine.
Nevertheless, accurately extracting SCR logic-level usage knowledge presents significant challenges, as it typically requires deep code understanding, which cannot be effectively achieved through simple keyword matching or template-based heuristics.
To address these challenges, we propose an LLM-powered agent, \aguke, for SCR analysis, which integrates an LLM with program analysis tools to extract reliable and semantically rich logic-level usage knowledge of SCRs.
Designing \aguke for SCR analysis involves overcoming three key challenges:
First, the LLMs integrated into \aguke lack inherent capabilities for autonomous task planning and step-wise reasoning, both of which are essential for managing complex extraction workflows.
Second, there is a lack of prior research or reference cases to guide the design of prompts specifically tailored for code comprehension and logic-level usage knowledge extraction, which is a critical aspect of \aguke.
Third, relying solely on generative LLMs for extracting low-level contract features proves unreliable due to their limited precision in program analysis.
We elaborate on our solutions to these challenges as follows.

\begin{figure}
\centering
\includegraphics[width=0.45\textwidth]{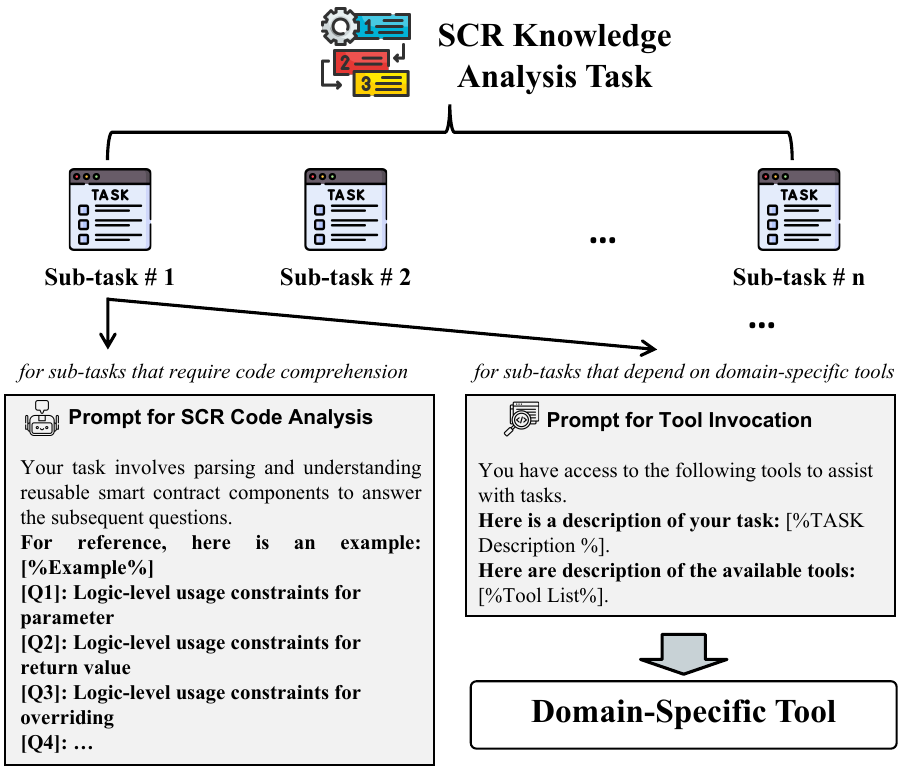}
\caption{The workflow of \aguke.}
\label{Figure: The workflow of aguke}
\end{figure}

\textbf{1) Task-Planning Middleware.}
To address the challenge of task planning, we design a middleware component that orchestrates the overall analysis workflow. 
As shown in Figure~\ref{Figure: The workflow of aguke}, the middleware decomposes each analysis task into a sequence of sub-tasks using a set of predefined rules.
Specifically, for sub-tasks that require code comprehension, the middleware directs \aguke to perform comprehension-oriented analysis. For sub-tasks that depend on domain-specific tools, the middleware instructs \aguke to resolve them by invoking the appropriate external tools. This modular task delegation mechanism ensures that each task is handled by the most suitable execution strategy, thereby improving both accuracy and efficiency in SCR analysis.

\textbf{2) Comprehension-Oriented Prompt.}
To address the challenge of knowledge extraction, we systematically analyze the potential causes of different types of logic-level usage violations of SCRs and formulate comprehension-oriented prompts to guide the LLM.
For instance, common sources of logic-level usage violations of SCRs include the absence of proper validation for input parameters passed to SCRs, and the failure to verify the return values of such calls for correctness or expected behavior.
Therefore, we design prompts that instruct \aguke to identify all implicit operations necessary for compliant usages of the given SCR. 

Figure~\ref{Figure: The workflow of aguke} illustrates the prompt template designed for \aguke. The questions included in the prompt are carefully crafted to support the extraction of logic-level usage knowledge of SCRs, and are effective for two primary reasons.
First, the questions are derived from a comprehensive analysis of hundreds of real-world cases of logic-level usage violations of SCRs, along with an in-depth examination of a large corpus of SCR source code. As such, they are highly relevant to common root causes of usage-related violation issues and collectively cover the vast majority of logic-level usage violation patterns.
Second, each question is formulated to elicit clear, concise, and structured responses from the LLM. This design ensures the outputs are machine-readable and directly usable in subsequent analysis stages.

\textbf{3) Domain-Specific Tool.}
To address the challenge of analyzing low-level features in smart contracts, we integrate a suite of specialized program analysis tools into \aguke. These tools enhance \aguke’s capabilities by enabling it to perform complex, domain-specific analysis tasks that are otherwise beyond the scope of LLMs alone.
One such tool is \textit{Slither}~\cite{feist2019slither}, a static analysis framework widely used for examining smart contracts. It supports a variety of program analysis techniques and provides rich semantic information. 
However, integrating \textit{Slither} directly into an LLM-based workflow presents practical obstacles. The LLM is not capable of directly invoking \textit{Slither}, and it also lacks the precision to interpret and filter the tool’s raw output effectively.
To bridge this gap, we develop a set of dedicated plugins that mediate between \aguke and the program analysis tools. These plugins are categorized into two functional types: tool-invoking plugins and response-processing plugins. The tool-invoking plugins encapsulate commonly used program analysis functionalities and expose them in a format that can be invoked by \aguke. The response-processing plugins reformat and restructure the outputs returned by these tools into a representation that is compatible with the \aguke's reasoning process. As summarized in Table~\ref{Table: Description for each plugin.}, each plugin is designed to ensure seamless integration and efficient interaction between program analysis tools and \aguke. 
Figure \ref{Figure: The workflow of aguke} illustrates the prompt designed for \aguke to invoke these plugins. The prompt begins by presenting a list of tools from which \aguke can make a selection. Once a tool is selected, it is executed via a tool-invoking plugin, and the results of tools are returned to \aguke through a response-processing plugin.

\subsection{RAG-Driven Inspector}
\label{RAG-driven Inspector}

\begin{figure}[ht]
\centering
\includegraphics[width=0.45\textwidth]{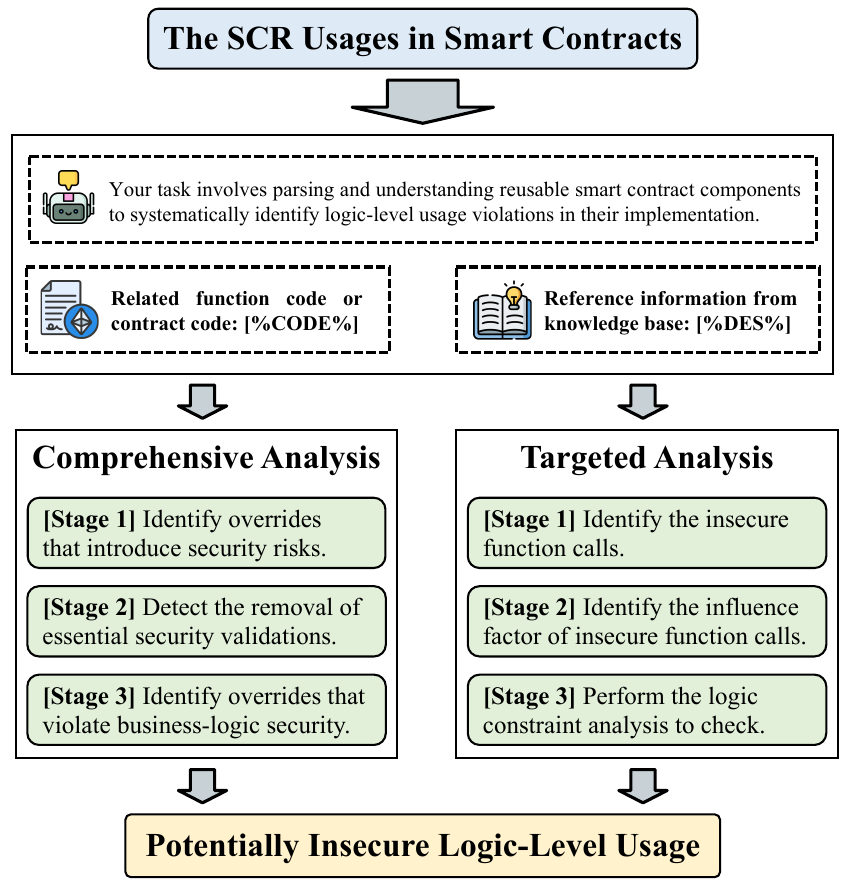}
\caption{The workflow of \agluc.}
\label{Figure: The workflow of agluc}
\end{figure}

While we have constructed a comprehensive SCR knowledge base, a practical and effective method for leveraging this knowledge in identifying potential insecure logic-level usage of SCRs remains absent. The challenge lies in the fact that SCRs often encapsulate highly business-specific semantics and exhibit complex usage patterns. Therefore, their correct usage requires deep semantic understanding and contextual reasoning, which cannot be achieved through simple rule-based matching.
To address the above challenges, we propose a RAG-driven inspector for identifying potential insecure logic-level usages of SCRs, which leverages the in-context learning capabilities of LLMs to perform nuanced behavior analysis. 
Nevertheless, implementing this approach requires the design of a well-structured and carefully organized workflow that facilitates seamless integration between retrieved knowledge and LLM-based reasoning.
In the following, we present our solutions to construct this RAG-driven inspector, \agluc, 
as shown in Figure~\ref{Figure: The workflow of agluc}.

\begin{figure}[ht]
\centering
\includegraphics[width=0.38\textwidth]{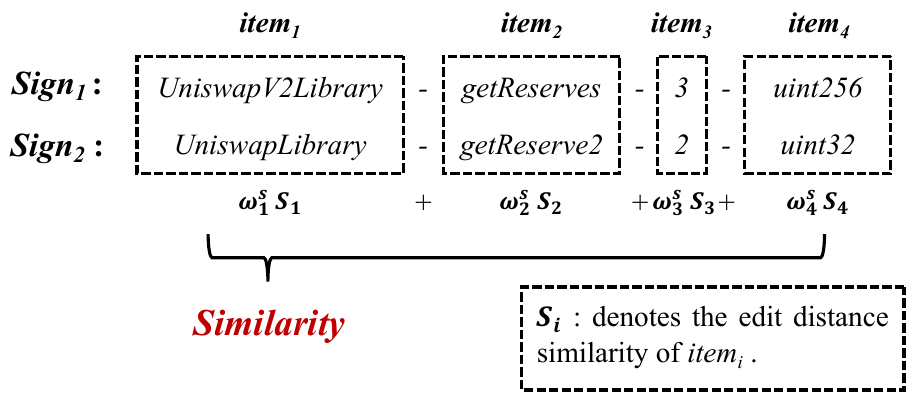}
\caption{Principle of similarity computation between the composite signature of the target SCR and those archived in the knowledge base.}
\label{Figure: The principles of similarity calculation}
\end{figure}

\noindent\textbf{Retrieval Strategy.}
As described in Section~\ref{SRC Knowledge Base Construction}, our knowledge base serves as a reference source to support the LLM during logic-level usage violation analysis. To effectively integrate this knowledge into \agluc, we design a retrieval strategy comprising the following steps. 
First, we generate composite signatures for all SCRs within the knowledge base. These signatures serve as the indexing keys for similarity-based retrieval. 
Next, we compute the similarity score between the composite signature of the target SCR and those stored in the knowledge base. As shown in Figure~\ref{Figure: The principles of similarity calculation}, the edit-distance similarity of each item in the composite signature is computed individually, and the overall similarity score is obtained as the weighted sum of the individual item similarities. 
The weighting coefficients for each item are determined through distribution-based statistical analysis, with $\boldsymbol{\omega}^{s} = (0.079,\, 0.421,\, 0.313,\, 0.187)$, as detailed in Appendix~\ref{Appendix: distributed-based}.
Finally, the SCR with the highest similarity score is selected as the reference candidate, providing contextual guidance for the LLM to assess logic-level usage violations of SCRs.

\noindent\textbf{Augmented Violation Analysis.}
In real-world auditing scenarios, auditors typically follow a systematic process when identifying logic-level usage violations of SCRs. They decompose the analysis of potentially insecure logic-level usages into two phases: comprehensive analysis and targeted analysis.
The comprehensive analysis focuses on the security risks introduced by SCR overriding and inheritance across the contract structure. During this phase, auditors assess whether the contract has removed security logic or introduced new logic that compromises security.
The targeted analysis, on the other hand, focuses on the security risks associated with SCR calls within the contract. In this phase, auditors begin by identifying potentially insecure SCR usages and then verifying their correctness and security relevance.
Inspired by this practical workflow, we design our prompt strategy to emulate the reasoning patterns commonly adopted by human auditors, thereby enabling the LLM to perform analogous stepwise analyses. 
Furthermore, we integrate a RAG mechanism into our prompts to provide relevant contextual information from the SCR knowledge base, thereby enhancing the model’s precision in assessing logic-level usage violations. 
This design also mirrors real-world auditing practices, where human auditors routinely consult extensive reference materials to make informed and accurate judgments.

Figure~\ref{Figure: The workflow of agluc} illustrates the prompt strategy designed for \agluc, which emulates the reasoning patterns commonly adopted by human auditors. The strategy consists of two phases: comprehensive analysis and targeted analysis. It first instructs the LLM to assume the role of a code auditor and then decomposes the reasoning process into six sequential stages.
In the comprehensive analysis phase (Stages C.1–C.3), the prompts guide the LLM to reason about SCR overriding and inheritance through a series of structured questions. The input for these stages includes the relevant smart contract code and reference information retrieved from the SCR knowledge base. Stage C.1 focuses on the newly added logic introduced through overriding and assesses whether such changes create new security risks. Stage C.2 examines the logic removed during overriding to determine whether essential protective measures have been eliminated. Stage C.3 inspects the modified business logic to evaluate whether the functional changes introduce potential security vulnerabilities. In the targeted analysis phase (Stages T.1–T.3), the prompts narrow the focus to the security implications of each SCR call within the contract. Stage T.1 takes as input the contract code, a list of extracted SCR calls, and a preliminary query to identify usages that may pose security risks. Stage T.2 introduces a follow-up prompt that guides the LLM to reason about how each identified SCR usage affects the overall contract security and to articulate its corresponding threat vectors. Stage T.3 enhances the model’s reasoning and verification by retrieving relevant reference information from the SCR knowledge base, informed by the previously identified usage patterns and threat vectors. This reference data supports a more accurate and explainable violation assessment through secondary validation.
Moreover, to ensure coherent reasoning and seamless interaction among stages and subsequent system modules, we design specific prompt rules for each stage. These rules constrain the model’s responses to structured formats, such as Yes/No, key–value pairs, or JSON, thereby facilitating consistent interpretation and integration throughout the analysis pipeline.

\subsection{Logic-Level Usage Violation Detection Engine}
\label{Noncompliance Detection Engine}

Though \system is capable of identifying potential insecure logic-level usages of SCRs through \agluc, we cannot fully rely on the LLM’s output due to its inherent randomness. In particular, even secure SCR usages may occasionally trigger false alerts. This issue arises primarily from the LLM’s hallucinations and sensitivity to similar SCR usage patterns, which can influence its judgment and lead to inconsistent assessments.
To address these challenges, we implement a logic-level usage violation detection engine that performs rigorous validation. This engine complements the LLM’s reasoning with additional layers of structural and context-sensitive verification, ensuring greater accuracy and robustness in logic-level usage violation detection. We craft a similarity-based checker and a snapshot-based inference conflict checker with the following design.

\noindent\textbf{Similarity-Based Checker.} 
In the analysis of SCR usages, both the internal structure and the associated logic-level usage features of an SCR are closely related to its compliance with logic-level usage requirements. To capture these factors, we design a similarity-based checker that computes the composite similarity between a target SCR usage instance and a reference instance retrieved from the knowledge base.
The similarity-based checker evaluates two key dimensions: the logical sequence and the latent structural embedding. 
In the logical sequence dimension, we measure the distance-based similarity of relevant logical constraints, denoted as $S_l$. 
In the latent structural embedding dimension, we compute the cosine similarity of numerical and control features, denoted as $S_n$ and $S_c$, respectively. 
The numerical features capture basic and PCG attributes, while the control features are derived from CFG attributes.
All relevant features are extracted using the multi-level feature extraction approach described in Section~\ref{Smart Contract Feature Extraction}.  
Finally, the principle of the similarity-based checker is defined in Equation~\ref{similarity:composite}, which identifies whether potentially insecure logic-level usages ($u$) satisfy the checking criteria. 
For potentially insecure logic-level usages identified through targeted analysis ($u \in G_T$), $S_n$, $S_c$, and $S_l$are considered. For those identified through comprehensive analysis ($u \in G_O$), only $S_n$ and $S_c$ are considered, since SCR overriding and inheritance do not involve external logical constraints as SCR calls do. A lower composite similarity score indicates a larger deviation of the SCR usage from the compliant usage patterns in the knowledge base. Specifically, lower values of $S_n$ and $S_c$ imply that the internal structure of the SCR has been inconsistently modified during its usage, whereas a lower $S_l$ suggests that fewer logical security constraints are enforced when invoking the SCR. 
\begin{IEEEeqnarray}{rCl}
\mathcal{S}(u)
&=&
\begin{cases}
\{u\mid \boldsymbol{\omega}^{\,o}\!\cdot(S_n(u),S_c(u)) < \tau_o\}, & u\in G_O, \\[8pt]
\{u\mid \boldsymbol{\omega}^{\,t}\!\cdot(S_n(u),S_c(u)) < \tau_t,\; \\
\quad S_l(u)<\tau_l\}, & u\in G_T. \\
\end{cases}
\IEEEeqnarraynumspace
\label{similarity:composite}
\end{IEEEeqnarray}
where 
$\boldsymbol{\omega}^{\,o}$ is $(0.27,0.73)$, $\boldsymbol{\omega}^{\,t}$ is $(0.42,0.58)$,  $\tau_o$ is $0.92$, $\tau_t$ is $0.68$, and $\tau_l$ is $0.90$. The weights and thresholds are empirically determined, and their soundness is verified through sensitivity analyses, as detailed in Appendix~\ref{Appendix: Sensitivity Analysis}.

\noindent\textbf{Snapshot-Based Inference Conflict Checker.}
LLMs exhibit strong capabilities in code comprehension and step-by-step security reasoning. 
Nevertheless, their propensity to generate hallucinated outputs, which are articulate yet incorrect responses, undermines the reliability of the system.
Specifically, when inference tasks exceed the LLM's capacity, it often generates plausible-sounding answers regardless of correctness, leading to a high rate of false positives.
To mitigate this issue, we propose a snapshot-based inference conflict checker that validates the consistency of intermediate reasoning steps throughout the analysis pipeline of \system.

\begin{figure}
\centering
\includegraphics[width=0.4\textwidth]{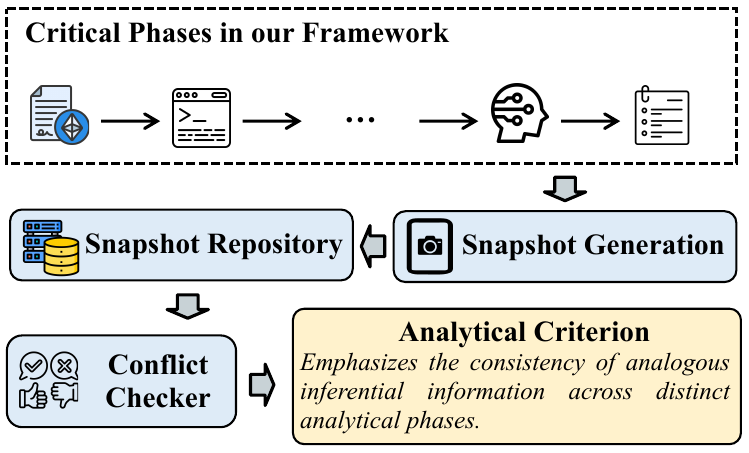}
\caption{The principles of snapshot-based inference conflict checker}
\label{Figure: Snapshot-based inference Conflict Checker}
\end{figure}

Figure~\ref{Figure: Snapshot-based inference Conflict Checker} illustrates the core principles of the snapshot-based inference conflict checker. The criterion of this checker is to ensure the consistency of analogous inferential information across distinct reasoning phases, thereby enhancing the overall trustworthiness of the analysis. We capture snapshots at critical phases of the \system workflow, each representing a summary of inferential outputs. Through empirical observation, we identify a set of analogous inference points across these phases, as summarized in Table~\ref{Table: Snapshot content across critical phases.}.
For instance, if our feature extraction identifies an SCR usage with no parameters, but \agluc incorrectly interprets it as an insecure parameter passing, this constitutes a conflict. We will treat such instances as LLM hallucinations and discard them.
These snapshots are stored in the snapshot repository, enabling systematic cross-phase consistency checks. The consistency analysis evaluates whether analogous inferential information, such as SCR signature, SCR security, or SCR definition, remains coherent across different analytical phases. If significant inconsistencies are detected, we infer that the task likely exceeds the LLM's reliable inference capacity and discard the result to avoid propagating hallucinated outputs.

\begin{table}[t]
\centering
\caption{Snapshot content across critical phases in \system.}
\label{Table: Snapshot content across critical phases.}
\resizebox{0.48\textwidth}{!}{
\begin{tabular}{lccccc}
\toprule[1.5pt]  
\multirow{2}{*}{ \textbf{Snapshot Content} } & 
\multirow{2}{*}{\textbf{Feature Extraction (\ref{Smart Contract Feature Extraction})}}   
& \multicolumn{4}{c}{\agluc \textbf{(\ref{RAG-driven Inspector})}} \\ \cmidrule{3-6}
& & \textbf{Stage 1} & \textbf{Stage 2} & \textbf{Stage 3} & \textbf{Stage 4}  \\ \midrule[1.5pt]
1) Signature of SCR       & \textcolor{green}{\ding{51}}  & \textcolor{green}{\ding{51}} & \textcolor{green}{\ding{51}}    & \textcolor{green}{\ding{51}}    & \textcolor{green}{\ding{51}}    \\  
2) Security of SCR         & \textcolor{red}{\ding{55}}  & \textcolor{green}{\ding{51}} & \textcolor{green}{\ding{51}}    & \textcolor{green}{\ding{51}}    & \textcolor{green}{\ding{51}}    \\  
3) Definition of SCR      & \textcolor{green}{\ding{51}}  & \textcolor{green}{\ding{51}} & \textcolor{red}{\ding{55}}    & \textcolor{green}{\ding{51}}    & \textcolor{red}{\ding{55}}    \\  
4) Parameters of SCR       & \textcolor{green}{\ding{51}} &  \textcolor{green}{\ding{51}} & \textcolor{red}{\ding{55}}    & \textcolor{red}{\ding{55}}    & \textcolor{green}{\ding{51}}    \\ 
5) Return Type of SCR     & \textcolor{green}{\ding{51}} & \textcolor{green}{\ding{51}}  & \textcolor{red}{\ding{55}}    & \textcolor{red}{\ding{55}}    & \textcolor{green}{\ding{51}}    \\ 
6) Parent Contract of SCR      & \textcolor{green}{\ding{51}} &  \textcolor{green}{\ding{51}} &  \textcolor{red}{\ding{55}}   & \textcolor{red}{\ding{55}}    & \textcolor{red}{\ding{55}}   \\ 
7) Overridden Function of SCR      & \textcolor{green}{\ding{51}} & \textcolor{green}{\ding{51}}  & \textcolor{red}{\ding{55}}    & \textcolor{red}{\ding{55}}    & \textcolor{red}{\ding{55}}    \\  
8) List of Related SCR calls    & \textcolor{green}{\ding{51}} & \textcolor{red}{\ding{55}}  & \textcolor{green}{\ding{51}}    & \textcolor{red}{\ding{55}}    & \textcolor{green}{\ding{51}}    \\ 
\bottomrule[1.5pt]
\end{tabular}
}
\begin{tablenotes}
\footnotesize
\item \textbf{Note:} \textcolor{green}{\ding{51}} indicates the information is captured at the corresponding phase, \textcolor{red}{\ding{55}} indicates absence.
\end{tablenotes}
\end{table}

\section{System Evaluation}
In this section, we evaluate the performance and capabilities of \system from multiple perspectives. Our evaluation is structured around the four key research questions:

\begin{itemize}[nosep, leftmargin=*]
\item[$\bullet$] \textbf{RQ1 (Knowledge Base Construction):}
How accurately can \system analyze SCR data and construct the associated logic-level usage knowledge base? (Section~\ref{RQ1})

\item[$\bullet$] \textbf{RQ2 (Violation Detection):}
How effective is \system in detecting logic-level usage violations of SCRs? (Section~\ref{RQ2})

\item[$\bullet$] \textbf{RQ3 (Ablation Study):}
How do the critical components of \system contribute to its overall detection performance? (Section~\ref{RQ3})

\item[$\bullet$] \textbf{RQ4 (Real-world Impact):}
How well does \system identify previously overlooked logic-level usage violations of SCRs in real-world on-chain smart contracts? (Section~\ref{RQ4})

\end{itemize}

Additionally, \textbf{DeepSeek-V3} serves as the foundational model for \system's evaluation.

\subsection{Datasets}

To enable our evaluation of \system's performance, we systematically collect $1,563$ smart contracts and $246$ SCRs from various sources. We create $3$ datasets.
The detailed description of these datasets is provided as follows.

\noindent\textbf{SCR Dataset ($D_{SCR}$).}
The $D_{SCR}$ dataset consists of $246$ SCRs with $7,422$ usage instances, collected from two sources as described in Section~\ref{SRC Knowledge Base Construction}. Specifically, among the $246$ SCRs, $40$ were selected based on usage frequency and $196$ were associated with historical logic-level usage violation incidents. These SCRs are chosen for knowledge base construction for two primary reasons: 1) they are widely adopted in real-world smart contracts, as demonstrated by prior research~\cite{icse/HuangCJZ24}; and 2) they exhibit a higher likelihood of inducing security risks, as evidenced by our analysis of historical attack statistics~\cite{defihacklabs}.

\noindent\textbf{Historical Violation Dataset ($D_{HV}$).}
The $D_{HV}$ dataset consists of smart contracts that have experienced attacks due to logic-level usage violations of SCRs. In particular, $D_{HV}$ includes $382$ smart contracts associated with $51$ logic-level usage violation cases, reported by BlockSec~\cite{blocksec}, DefiHackLabs~\cite{defihacklabs}. These real-world attack instances serve as high-fidelity benchmarks to evaluate the practical effectiveness of \system in detecting logic-level usage violations of SCRs.

\noindent\textbf{Real-World On-Chain Dataset ($D_{RO}$).}
The $D_{RO}$ dataset comprises smart contracts deployed on prominent blockchains and sourced from leading audit platforms, such as Code4rena~\cite{Code4rena}. It includes $1,181$ contracts collected from these sources. These contracts represent modern development trends, incorporating contemporary business models and programming paradigms, thus providing a representative testbed for assessing the generalizability of \system.

\subsection{Evaluation of Knowledge Base Construction}
\label{RQ1}

In this step, we evaluate the construction quality of the SCR knowledge base using precision as the primary metric. Within \system, we design an SCR analysis agent, \aguke, to facilitate automated and accurate knowledge extraction for SCR logic-level usage. We begin by assessing the precision of the SCR knowledge extracted by \aguke, which is foundational for downstream logic-level usage violation detection. To further understand the design efficacy of \aguke, we conduct an ablation study. The critical design principle of \aguke lies in its integration of a suite of specialized program analysis tools with LLM, thereby enhancing \aguke's analytical precision and contextual understanding. For the ablation, we compare two versions of the agent: the full version, \aguke, and a variant without integrated tools, \aguke~\textit{w/o Tools}. 
We conduct the above evaluation on $D_{SCR}$.

To evaluate the performance of \aguke on $D_{SCR}$, we obtain a total of $7,422$ analysis results. 
Given the difficulty of manually verifying each result at scale, we conduct a sampling-based evaluation to assess correctness and reliability.
To enhance the credibility of our evaluation, we adopt three complementary sampling strategies from different perspectives: complexity-based, frequency-based, and random-based sampling.
\textbf{First}, to assess \aguke’s capability in analyzing complex SCRs, we employ a complexity-based strategy. SCRs in $D_{SCR}$ exhibit varying levels of complexity, ranging from simple components comprising only 1–2 lines of code, to more complex components involving multiple statements, external calls, or nested invocations. 
We select SCRs that contain more than four function calls, resulting in a total of $210$ SCR usage instances.
\textbf{Second}, to evaluate \aguke’s capability in handling SCR with similar signatures but variations in internal structures, we adopt a frequency-based sampling strategy. 
Specifically, we group SCRs by function name and select those groups that contain more than six distinct usage variations, resulting in a total of $212$ SCR usage instances.
\textbf{Third}, we apply a random-based strategy by randomly selecting $222$ SCR usage instances from $D_{SCR}$. This ensures an unbiased and diverse coverage across different usage scenarios, enhancing the generalizability of our evaluation.

\begin{table}[t]
\caption{Precision of \aguke in SCR analysis.}
\centering
\label{Table: Precision of uke}
\resizebox{0.38\textwidth}{!}{
\begin{tabular}{c|c|c|c}
\toprule[1.5pt]
\textbf{Strategy} &  \textbf{TP} & \textbf{FP} & \textbf{Precision} \\ \midrule[1.5pt]
Complexity-Based Sampling     & 192      & 18      & 91.43\%             \\ 
Frequency-Based Sampling      &  202      &  10     &  95.28\%    \\
Random-Based Sampling          & 209       & 13      & 94.14\%              \\
\midrule[0.5pt]
\textbf{Overall}  & \textbf{603} & \textbf{41}   & \textbf{93.63\%} \\
\bottomrule[1.5pt]
\end{tabular}
}
\end{table}

\noindent\textbf{Precision Evaluation.}
As shown in Table~\ref{Table: Precision of uke}, \aguke achieves high precision in analyzing SCR usage, with an overall precision of $93.63\%$. Specifically, it attains a precision of $91.43\%$ for SCR usage instances selected via the complexity-based strategy, $95.28\%$ via the frequency-based strategy, and $94.14\%$ via the random-based strategy.
Further analysis of the false positives reveals that \aguke occasionally extracts incorrect logic-level usage patterns for certain SCRs. For instance, when the input parameter involves an Ethereum address, \aguke tends to output “non-zero address”, which represents a syntax-level rule rather than a logic-level usage constraint. In contrast, for input parameters involving token amounts, \aguke generates recommended value ranges and their logic relationships with user token balances, which appropriately reflect logic-level usage.

\noindent\textbf{Tool Integration Impact.}
As shown in Table~\ref{Table: Comparison of uke and uke wo tools.}, \aguke demonstrates substantial performance improvements over \aguke~\textit{w/o Tools}. Specifically, \aguke achieves a precision of $93.63\%$, whereas \aguke~\textit{w/o Tools} yields a lower precision of $87.27\%$, due to both a reduction in true positives and an increase in false positives. The integration of program analysis tools enables \aguke to significantly enhance the precision of SCR usage extraction by improving the identification of valid patterns and reducing incorrect inferences. 
We further analyze the false positives by \aguke~\textit{w/o Tools}. The primary issue arises from its difficulty in correctly interpreting low-level features of SCR code. 
For instance, when instructed to infer the logic-level secure range of input parameters, some false positives from \aguke~\textit{w/o Tools} can be attributed to its misinterpretation of low-level code features. Specifically, in some SCRs with no or very few input parameters, the agent misidentifies locally defined variables as input parameters. This behavior stems from its limited ability to distinguish between actual input parameters and internal variables. In contrast, the tool-integrated version of \aguke accurately identifies these low-level features, thereby avoiding such misinterpretations.

\begin{table}[t]
\centering
\caption{Comparison of \aguke and \aguke~\textit{w/o Tools}.}
\label{Table: Comparison of uke and uke wo tools.}
\resizebox{0.45\textwidth}{!}{
\begin{tabular}{c|c|c}
\toprule[1.5pt]
\multirow{2}{*}{\textbf{Strategy}} & \multicolumn{2}{c}{\textbf{Precision}}    \\ 
                                & \aguke  & \aguke~\textit{w/o Tools} 
                               \\ \midrule[1.5pt]
Complexity-Based Sampling          & 91.43\%                & 82.86\%             \\ 
Frequency-Based Sampling        & 95.28\%                & 90.57\%             \\ 
Random-Based Sampling & 94.14\%                      & 88.29\%              \\
\midrule[0.5pt]
\textbf{Overall}  &  \textbf{93.63\%} & \textbf{87.27\%}  \\
\bottomrule[1.5pt]
\end{tabular}
}
\end{table}

\begin{takeaway}
    \textbf{RQ1:}  \aguke in \system exhibits strong performance in SCR analysis, achieving a precision of $93.63\%$. Moreover, our tool integration strategy enhances the \aguke's precision by $6.36\%$.
\end{takeaway}

\subsection{Evaluation of Violation Detection}
\label{RQ2}

This subsection addresses RQ2. We evaluate the effectiveness of \system in detecting logic-level usage violations of SCRs using three standard metrics: Precision, Recall, and F1-score. To the best of our knowledge, we are the first to address logic-level usage violations of SCRs. Though previous smart contract vulnerability detectors are not specifically designed for this purpose, some vulnerabilities reported by these detectors stem from the logic-level usage violations of SCRs. Therefore, we compare \system against representative vulnerability detectors. Specifically, we select five widely used detectors: \textit{Smartinv}~\cite{sp/smartinv2024}, \textit{Smartian}~\cite{choi2021SMARTIAN}, \textit{Mythril}~\cite{mythril}, \textit{Falcon}~\cite{falcon}, and \textit{Slither}~\cite{feist2019slither}. 
\textit{Smartinv} is an LLM-powered verification framework that leverages LLMs to identify the vulnerability and infer security invariants. 
Comparing \system with LLM-based approaches such as \textit{SmartInv} not only allows for evaluating its performance advantages but also helps mitigate potential risks of LLM-induced data leakage.
\textit{Smartian} is a grey-box fuzzer that utilizes data-flow analysis to guide input generation. \textit{Mythril} is a symbolic execution engine tailored for smart contract vulnerability detection. Both \textit{Falcon} and \textit{Slither} are static analysis tools that detect vulnerabilities based on a large corpus of hand-crafted rules. 
For a fair comparison, we only consider the subset of vulnerabilities reported by these tools that are attributable to logic-level usage violations of SCRs.
We evaluate all tools on $D_{HV}$.

\begin{table}[t]
\centering
\caption{Logic-level usage violation detection accuracy.}
\label{Table: noncompliance detection accuracy.}
\resizebox{0.45\textwidth}{!}{
\begin{tabular}{c|c|c|c|c|c|c}
\toprule[1.5pt]
\multirow{2}{*}{\textbf{Tool}} & \multicolumn{6}{c}{\textbf{$D_{HV}$}}    \\ 
                               & \textbf{TP} & \textbf{FP} & \textbf{FN}  & \textbf{Precision} &\textbf{Recall} & \textbf{F1-Score} \\
                                \midrule[1.5pt]
\system           & \textbf{42} & \textbf{10} & \textbf{9} & \textbf{80.77\%}    & \textbf{82.35\%}   & \textbf{81.55\%}       \\ 
\textit{Smartinv}      & 19 & 7 & 32 & 73.08\%  & 37.25\%  & 49.35\%  \\
\textit{Smartian}   &  3 & 0 & 48 & 100.00\% & 5.88\%  &  11.11\%    \\
\textit{Mythril}     & 2 & 1 & 49      & 66.67\% & 3.92\%  & 7.41\%      \\ 
\textit{Falcon}      & 18 & 59 & 33   & 23.38\%   & 35.29\%  & 28.13\%  \\
\textit{Slither}     & 16 & 65 & 35     & 19.75\%  & 31.38\%  & 24.24\%      \\ 
\bottomrule[1.5pt]
\end{tabular}
}
\end{table}

As shown in Table~\ref{Table: noncompliance detection accuracy.}, \system outperforms all baselines in detecting logic-level usage violations of SCRs. It achieves $42$ true positives, $10$ false positives, and $9$ false negatives, corresponding to a precision of $80.77\%$, a recall of $82.35\%$, and an F1-score of $81.55\%$.
We further analyze the causes of false positives and false negatives in \system. 
First, imprecise SCR knowledge extracted by \aguke constitutes the primary cause of both false positives and false negatives, as perfect extraction of logic-level usage semantics remains inherently challenging.
Second, inaccuracies in the analysis performed by \agluc may also contribute to certain false positives and false negatives.

In addition, we further investigate the false positives and false negatives produced by the baselines. For \textit{Smartinv}, its primary limitation lies in the inability to generate accurate and verifiable invariants necessary for effective detection. \textit{Smartian} and \textit{Mythril} struggle to simulate the execution paths that lead to logic-level usage violations, and their predefined detection patterns are often insufficient for capturing such issues. 
\textit{Falcon} and \textit{Slither} employ a large set of detection rules derived from historical vulnerability patterns. While this rule-based approach enables them to identify a greater number of true positives compared to other baselines, it also leads to a high incidence of false positives due to the limited contextual understanding and coarse granularity of these predefined rules.

\begin{takeaway}
    \textbf{RQ2:}  \system demonstrates strong performance in detecting logic-level usage violations of SCRs, achieving a precision of $80.77\%$, a recall of $82.35\%$, and an F1-score of $81.55\%$.
\end{takeaway}

\subsection{Ablation Study}
\label{RQ3}

This subsection addresses RQ3. We conduct the ablation study from two perspectives. 
\textbf{First,} to evaluate the contribution of critical modules to violation detection accuracy, we evaluate three variants of \system:
1)~\system~\textit{w/o}~\aguke, 2)~\system~\textit{w/o}~\agluc, and 3)~\system~\textit{w/o~Sim Checker}. Specifically, \system~\textit{w/o}~\aguke disables the SCR knowledge base constructed by \aguke~(Section~\ref{SRC Knowledge Base Construction}).
\system~\textit{w/o}~\agluc disables the RAG-driven inspector powered by \agluc~(Section~\ref{RAG-driven Inspector}). And \system~\textit{w/o Sim Checker} disables the similarity-based checker(Section~\ref{Noncompliance Detection Engine}).
We perform the above variants in the dataset $D_{HV}$.
\textbf{Second,} to evaluate the role of the snapshot-based inference conflict analysis in mitigating LLM hallucinations, we perform an in-depth analysis of the outputs generated by \agluc.

\noindent\textbf{Impact on Detection Accuracy.}
As shown in Table~\ref{Table: Comparison of three variants.}, \system consistently outperforms all ablated variants.
In particular, \system~\textit{w/o}~\aguke achieves a precision of $55.77\%$, a recall of $56.86\%$, and an F1-score of $56.31\%$, indicating that the absence of a structured knowledge base significantly degrades performance.
\system~\textit{w/o}~\agluc achieves a precision of $30.07\%$, a recall of $84.31\%$, and an F1-score of $44.33\%$, underscoring the importance of the RAG-driven inspector in balancing precision and recall.
\system~\textit{w/o~Sim Checker} achieves a precision of $47.25\%$, a recall of $84.31\%$, and an F1-score of $60.56\%$, demonstrating the necessity of reducing false positives.

\begin{table}[t]
\centering
\caption{Comparison of \system and three variants.}
\label{Table: Comparison of three variants.}
\resizebox{0.45\textwidth}{!}{
\begin{tabular}{c|c|c|c}
\toprule[1.5pt]
\multirow{2}{*}{\textbf{Variant}} & \multicolumn{3}{c}{\textbf{$D_{HV}$}}    \\ 
                                & \textbf{Precision}  & \textbf{Recall} & \textbf{F1-Score} \\
                                \midrule[1.5pt]
\system      & \textbf{80.77\%}    & \textbf{82.35\%}  & \textbf{81.55\%}       \\ 
\system~\textit{w/o}~\aguke           & 55.77\%    & 56.86\%  & 56.31\%      \\ 
\system~\textit{w/o}~\agluc           & 30.07\%    & 84.31\%  & 44.33\%  \\
\system~\textit{w/o~Sim Checker}       & 47.25\%    & 84.31\%  & 60.56\%  \\
\bottomrule[1.5pt]
\end{tabular}
}
\end{table}

\noindent\textbf{Effectiveness in Mitigating Hallucinated Outputs.}
As shown in Table~\ref{Table: Effectiveness of Snapshot-Based Inference Conflict Checker}, \agluc initially produces $2,684$ candidate outputs.
The snapshot-based inference conflict checker identifies $1,957$ outputs ($72.91\%$) as hallucinated due to the inconsistencies among their snapshots.
Consequently, only $727$ outputs are retained for further analysis.
This high filtering rate confirms the checker’s effectiveness in suppressing unreliable \agluc-generated information, thereby enhancing the overall robustness and reliability of \system.

\begin{table}[t]
\centering
\caption{Effectiveness of snapshot-based inference conflict checker in reducing \agluc-generated hallucinations.}
\label{Table: Effectiveness of Snapshot-Based Inference Conflict Checker}
\resizebox{0.35\textwidth}{!}{
\begin{tabular}{ccc}
\toprule[1.5pt]
\textbf{Processing Stage} &  \textbf{Count} & \textbf{Percentage} \\
\midrule[1.5pt]
Initial Generation  & 2,684 & 100.00\% \\
Conflicts Identified  & 1,957 & 72.91\% \\
Final Validated Entries  & 727 & 27.09\% \\
\bottomrule[1.5pt]
\end{tabular}
}
\end{table}

\begin{takeaway}
    \textbf{RQ3:}  The integration of \aguke, \agluc, and a similarity-based checker significantly enhances the violation detection performance of \system. Furthermore, the snapshot-based inference conflict checker effectively mitigates the LLM's hallucinations, substantially enhancing the robustness of \system.
\end{takeaway}

\subsection{Real-world Impact}
\label{RQ4}

This subsection addresses RQ4. We deploy \system on the real-world on-chain dataset $D_{RO}$ to assess its effectiveness in detecting logic-level usage violations of SCRs, including the identification of zero-day vulnerabilities resulting from logic-level usage violations in real-world on-chain smart contracts. 

\system successfully identified $13$ zero-day vulnerabilities within the above smart contracts. As of now, $9$ of these zero vulnerabilities have been assigned CVE IDs. These findings have been responsibly disclosed to the respective vendors. Notably, one of the most prominent DeFi vendors, anonymized as \textit{AnonymDeFi}, has acknowledged our report. According to their security team, the identified vulnerability poses a severe threat to their platform’s financial security. \textit{AnonymDeFi} exhibits high liquidity, with a \textbf{24-hour Trading Volume-to-Market Capitalization (Vol/Mkt Cap)} ratio ranging from \textbf{$0.13$ to $0.18$}. Specifically, its \textbf{market capitalization exceeds $\$1$ billion}, and its \textbf{24-hour Trading Volume surpasses $\$150$ million}. Therefore, any vulnerabilities within its contracts could lead to significant economic consequences. Due to the sensitive nature of these zero-day vulnerabilities and the fact that they remain unpatched at the time of writing, we anonymize all identifying details. A representative case study, with critical information redacted for confidentiality, is provided below.

\begin{figure}[t]
\begin{lstlisting}[
      style=soltight,
      firstline=1, lastline=1,
      firstnumber=1
    ]
contract SimpleToken is ERC20Capped, Ownable {
\end{lstlisting}
    {\begingroup
    \lstset{backgroundcolor=\color{hl}, firstnumber=last}
\begin{lstlisting}[
      style=soltight,
      firstline=2, lastline=3
    ]
  constructor(uint256 _initialSupply, address initialOwner) ERC20("SimpleToken", "Simple") ERC20Capped(1000000000 * 10 ** 18) Ownable(initialOwner) 
  { ERC20._mint(msg.sender, _initialSupply); } }
\end{lstlisting}
    \endgroup}
\begin{lstlisting}[
      style=soltight,
      firstline=4,
      firstnumber=last
    ]
contract ERC20{
  function _mint(address account, uint256 value) internal {
    if (account == address(0)) { revert ERC20InvalidReceiver(address(0)); }
    _update(address(0), account, value); }
  function _update(address from, address to, uint256 value) internal virtual { ... } }
contract ERC20Capped is ERC20{
  uint256 private immutable _cap;
  function _update(address from, address to, uint256 value) internal virtual override {
    super._update(from, to, value);
    if (from == address(0)) {
      uint256 maxSupply = cap();  
      uint256 supply = totalSupply(); 
      if (supply > maxSupply) { 
        revert ERC20ExceededCap(supply, maxSupply); 
      } } } }
\end{lstlisting}
    \caption{Bypassing the supply cap enables arbitrary token minting (lines 2-3).}
    \label{Figure: Case Study}
\end{figure}

As shown in Figure~\ref{Figure: Case Study}, the code exhibits a logic-level usage violation of SCRs, allowing the sender to mint tokens without enforcing the intended supply cap. 
Specifically, the \solcode{SimpleToken} contract inherits from \solcode{ERC20Capped}, an SCR designed to impose a maximum token supply of \solcode{1e27}  at line 2. Subsequently, the contract invokes \solcode{ERC20.\_mint(msg.sender, \_initialSupply)} at line 3 to mint the initial token supply. However, this invocation ultimately calls \solcode{ERC20.\_update} at line 7 to update the total supply, where no check against the \solcode{cap} is performed. 
As a result, the supply cap enforcement is silently bypassed, allowing the sender to mint tokens arbitrarily, which poses significant financial risks to other users.
To address logic-level usage violations of \solcode{ERC20Capped}, the correct usage requires overriding the \solcode{\_mint} function in \solcode{SimpleToken} to delegate minting to \solcode{ERC20Capped.\_update}, thereby enforcing the required supply cap verification (lines 14–17).

\begin{takeaway}
    \textbf{RQ4:} \system successfully detects $13$ zero-day vulnerabilities in real-world on-chain smart contracts, $9$ of which have been assigned CVE IDs. Notably, the zero-day vulnerabilities identified in one of the most prominent DeFi platforms, with a market capitalization exceeding \$1 billion and a 24-hour trading volume surpassing \$150 million, have been confirmed by their security team.
\end{takeaway}

\section{Discussion}

Although \system demonstrates strong performance in detecting logic-level usage violations of SCRs, it still has several limitations. First, the presence of false positives and false negatives is primarily attributed to imprecise SCR knowledge extraction and occasional failures in the RAG-driven analysis. While \aguke achieves high precision in extracting SCR knowledge, it struggles with particularly complex cases. 
Similarly, \agluc currently focuses on analyzing parameter and return-value constraints, and reasoning about the logical differences arising from inheritance and overriding, which may still overlook certain exceptional cases.
Future work will focus on enhancing both \aguke and \agluc by integrating additional program analysis tools and developing more sophisticated reasoning strategies.
Second, the inherent randomness and hallucination tendencies of LLMs can undermine the overall reliability of \system. Although the snapshot-based inference conflict checker effectively mitigates these issues, it does not fully eliminate them. As a future direction, we plan to fine-tune a dedicated LLM tailored to logic-level usage violation detection, aiming to further improve reasoning accuracy and system robustness.

\section{Related work}

\noindent\textbf{Smart Contract Analysis.}
Many works have been conducted on smart contract security, primarily through static and dynamic analysis techniques. Static analysis methods examine source code or bytecode to capture deep structural or semantic features for vulnerability detection. Representative tools include \textit{Slither}~\cite{feist2019slither}, \textit{AutoAR}~\cite{ndss/AutoAR2025}, \textit{Securify}~\cite{securify_ccs_2018}, \textit{Zeus}~\cite{ndss/zeus2018}, and \textit{MadMax}~\cite{Madmax_oopsla_2018}. Dynamic analysis approaches assess contract behavior through runtime testing. Notable tools include  \textit{Smartian}\cite{choi2021SMARTIAN}, \textit{ItyFuzz}~\cite{shou2023ityfuzz}, \textit{Harvey}~\cite{harvey_FSE_2020}, \textit{ContractFuzzer}~\cite{contractfuzzer_ase_2018}, and \textit{Echidna}~\cite{echidna_issta_2020}. Currently, LLMs have also been applied to smart contract analysis, such as \textit{PROMFUZZ}~\cite{ASE25/promfuzz}, \textit{PropertyGPT}~\cite{ndss/propertygpt2025}, \textit{SmartInv}~\cite{sp/smartinv2024}, and \textit{GPTScan}~\cite{icse/gptscan2024}. For instance, \textit{PropertyGPT} leverages LLMs to automatically generate verification properties for unknown code, enabling comprehensive formal analysis. Nevertheless, applying these methods in our task is not trivial since they are not designed to detect logic-level usage violations of SCRs.

\noindent\textbf{Reusable Components Analysis.} 
Numerous studies have been devoted to detecting security issues arising from component reuse across a wide range of software systems. These approaches can be broadly categorized into two groups based on their analytical perspectives. The first category aims to assess the internal security of reusable components themselves, such as \textit{Vulture}~\cite{ndss/Vulture2025}, \textit{VAScanner}~\cite{tse/VAScanner24}, \textit{ATVHUNTER}~\cite{icse/ATVHUNTER2021}, \textit{LibID}~\cite{issta/LibID19}, and \textit{LibScout}~\cite{ccs/LibScout2016}. The second category focuses on detecting the misuse of reusable components during integration, such as \textit{APP-Miner}~\cite{sp/APPMiner2025},\textit{GPTAid}~\cite{ndss/GPTAid2025}, \textit{ARBITRAR}~\cite{sp/ARBITRAR2021}, \textit{Advance}~\cite{ccs/Advance2020}, \textit{MutApi}~\cite{icse/MutApi2019}, \textit{APIScan}~\cite{uss/APIScan2016} and \textit{APEx}~\cite{ase/APEx2016}. However, these detection approaches primarily target misuse patterns and internal security issues within their respective software environments. Even when adapted to smart contracts, they remain ineffective at identifying logic-level usage violations of SCRs. In addition, several studies have investigated the security of reusable components in smart contracts, such as \textit{RHT}~\cite{icse/HuangCJZ24}, \textit{EquivGuard}~\cite{issta/equivguard2025}, and \textit{ZepScope}~\cite{uss/zepscope2025}. For example, \textit{RHT} introduces eight patterns of library misuse to assist developers in avoiding common pitfalls and mitigating potential financial losses. \textit{ZepScope} presents a systematic analysis of the security posture of the widely used \textit{OpenZeppelin} library in real-world contracts. \textit{EquivGuard} leverages a combination of static taint analysis and symbolic execution to identify six categories of EVM-inequivalence issues in cross-chain contract reuse. Although these methods offer valuable insights into the SCR security, they overlook the analysis of the logic-level usage feature of such components, thereby failing to detect logic-level usage violations of SCRs.

\section{Conclusion}
In this paper, we introduce \system, the first automated and practical system for detecting logic-level usage violations of SCRs, which bridges a critical gap in existing smart contract research.
\system achieves a precision of $80.77\%$, a recall of $82.35\%$, and an F1-score of $81.55\%$ in detecting logic-level usage violations of SCRs, representing a significant performance improvement over state-of-the-art vulnerability detectors. Additionally, we have created the first ground-truth dataset for logic-level usage violations of SCRs, including $382$ contracts from various DeFi projects. Furthermore, our analysis of $1,181$ real-world contracts has identified $13$ zero-day vulnerabilities, with $9$ of which have been assigned CVE IDs. The identified vulnerabilities have been confirmed by one of the most prominent DeFi platforms with a market capitalization exceeding \$1 billion and a 24-hour trading volume surpassing \$150 million.

\section*{Ethics Considerations}
We pay careful attention to potential ethical concerns associated with this work. First, all smart contracts analyzed in our study are sourced from publicly available blockchain platforms or reputable auditing platforms, ensuring transparency and legitimacy. Second, all proof-of-concept attacks are conducted exclusively in a controlled local environment, without interacting with deployed contracts, thereby avoiding any real-world consequences. Finally, we have responsibly disclosed all identified bugs to the corresponding vendors to support the improvement of their security practices.
\section*{LLM Usage Considerations}

In this work, we strictly follow the policy on the responsible use of LLMs, as detailed below. 

\noindent\textbf{Originality.} LLMs are used to improve the clarity and fluency of writing. All research content, experimental design, and conclusions are authored, reviewed, and verified entirely by the authors. We therefore take full responsibility for the originality of this paper.

\noindent\textbf{Transparency and Responsibility.} LLMs are employed in the SCR Knowledge Base Construction and RAG-Driven Inspector modules to assist with information extraction and reasoning. 
To ensure transparency, reproducibility, and responsible usage, we summarize our practices as follows. 
First, we adopt the open-source model DeepSeek-V3 without performing any fine-tuning, thereby ensuring the transparency and integrity of LLM usage. Second, the raw data used for SCR knowledge construction are collected entirely from publicly accessible blockchain platforms, ensuring that our analysis is based on open and verifiable sources. Third, all processes and configurations involving LLM usage are explicitly documented in the paper to ensure experimental reproducibility.

\bibliographystyle{IEEEtran}

\bibliography{Section/Ref}

\newpage

\appendices

\section{Distribution-based Weight Inference for Composite Signature Retrieval}
\label{Appendix: distributed-based}

Each SCR composite signature is represented as a four-dimensional tuple 
$Sign = (feature_1, feature_2, feature_3, feature_4)$, 
where $feature_1$ and $feature_2$ correspond to the \textit{contract name} and \textit{function name}, 
and $feature_3$ and $feature_4$ correspond to the \textit{number of parameters} and \textit{return value}, respectively. 
The first two features are continuous in lexical semantics, while the latter two are discrete. 
Accordingly, we assign equal group priors, 
$W_{\mathcal{C}} = 0.5$ and $W_{\mathcal{D}} = 0.5$, 
where $\mathcal{C} = \{1,2\}$ and $\mathcal{D} = \{3,4\}$ 
denote the continuous and discrete feature groups.

To capture the inherent stability of each feature, 
we infer its weight from the empirical distribution of string lengths 
collected from the knowledge base. 
For each feature $i$, the observed length samples are denoted as 
$\boldsymbol{\ell}_i = \{\ell_i^{(1)}, \ell_i^{(2)}, \dots, \ell_i^{(N)}\}$. 
Each sample is normalized by its total length, as defined in Equation~\ref{eq:length-norm}.

\begin{IEEEeqnarray}{rCl}
    \hat{\ell}_i^{(n)} = \frac{\ell_i^{(n)}}{\sum_{m=1}^{N}\ell_i^{(m)}}, 
    \quad n = 1,\dots,N.
    \label{eq:length-norm}
\end{IEEEeqnarray}

The sample variance of the normalized lengths is then computed 
using Equation~\ref{eq:variance}, which measures the dispersion of the length distribution:
\begin{IEEEeqnarray}{rCl}
s_i^2 = \frac{1}{N-1}\sum_{n=1}^{N}
\bigl(\hat{\ell}_i^{(n)} - \overline{\hat{\ell}}_i\bigr)^2.
\label{eq:variance}
\end{IEEEeqnarray}

To prevent numerical instability when a feature exhibits extremely low variance, 
we introduce a small stability constant $\epsilon$, as shown in Equation~\ref{eq:epsilon}.
\begin{IEEEeqnarray}{rCl}
\epsilon = 0.01 \cdot 
\mathrm{Median}\bigl(\{s_j^2\}_{j\in\mathrm{group}(i)}\bigr).
\label{eq:epsilon}
\end{IEEEeqnarray}

The intra-group weight $\tilde{w}_i$ of each feature is then 
determined through inverse-variance normalization, 
as defined in Equation~\ref{eq:intra-group}. 
This operation assigns larger weights to features with more stable length distributions.
\begin{IEEEeqnarray}{rCl}
\tilde{w}_i = 
\frac{(s_i^2+\epsilon)^{-1}}
{\sum_{j\in\mathrm{group}(i)} (s_j^2+\epsilon)^{-1}}.
\label{eq:intra-group}
\end{IEEEeqnarray}

Finally, the global weight $w_i$ is obtained 
by combining the group priors with the intra-group coefficients, 
as described in Equation~\ref{eq:global-weight}.
\begin{IEEEeqnarray}{rCl}
\omega^{s}_i =
\begin{cases}
W_{\mathcal{C}} \cdot \tilde{w}_i, & i \in \mathcal{C}, \\[3pt]
W_{\mathcal{D}} \cdot \tilde{w}_i, & i \in \mathcal{D}.
\end{cases}
\label{eq:global-weight}
\end{IEEEeqnarray}

In our empirical evaluation, the resulting weights are 
$\boldsymbol{\omega}^{s} = (0.079,\, 0.421,\, 0.313,\, 0.187)$.

\section{Sensitivity Analysis of Weights and Thresholds in the Similarity-Based Checker}
\label{Appendix: Sensitivity Analysis}

To evaluate the soundness of both weights and the thresholds in the similarity-based checker, we conducted a sensitivity analysis.

For each group of weights $\boldsymbol{\omega} = (\omega_n, \omega_c)$ satisfying $\omega_n + \omega_c = 1$, we varied $\omega_n$ within the interval $[\omega_n - 0.05,\, \omega_n + 0.05]$ with a step size of $0.01$, while $\omega_c$ is adjusted accordingly to maintain the normalization constraint. This setting provides a sufficiently fine granularity to capture parameter sensitivity without overfitting to small fluctuations, which is appropriate for discrete-task evaluation where performance metrics change non-continuously. For each threshold $\tau$, we varied its value within $\pm5\%$ of its original setting with a step of $1\%$. This range ensures a reasonable local perturbation that reflects potential deviations caused by empirical estimation errors or environmental variance, while keeping the perturbation narrow enough to preserve the detection semantics of the model. Table~\ref{Table: weight and threshold.} provides the detailed configuration of each parameter.

\begin{table}[h]
\centering
\caption{The configuration of weights and thresholds in sensitivity analysis.}
\label{Table: weight and threshold.}
\resizebox{0.48\textwidth}{!}{
\begin{tabular}{c|c|c|c|c}
\toprule[1.5pt]
\multicolumn{2}{c|}{\textbf{Parameter}} & \textbf{Value} & \textbf{Interval}  &  \textbf{Step}\\
\midrule[1.5pt]
\multirow{2}{*}{$\boldsymbol{\omega}^{t}$} & $\omega^{t}_{n}$  & 0.42 & $[\omega^{t}_{n} - 0.05, \, \omega^{t}_{n} + 0.05]$ & 0.01   \\ \cmidrule{2-5}
 & $\omega^{t}_{c} = 1 - \omega^{t}_{n}$ & 0.58 & $[ \, \omega^{t}_{c} + 0.05, \, \omega^{t}_{c} - 0.05 \,]$ & 0.01   \\ \midrule
\multirow{2}{*}{$\boldsymbol{\omega}^{o}$} & $\omega^{o}_{n}$ & 0.27 & $[ \, \omega^{o}_{n} - 0.05, \, \omega^{o}_{n} + 0.05 \, ]$ & 0.01   \\ \cmidrule{2-5}
 & $\omega^{o}_{c} = 1 - \omega^{o}_{n}$ & 0.73 &  $[ \, \omega^{o}_{c} + 0.05, \, \omega^{o}_{c} - 0.05 \, ]$ & 0.01   \\  \midrule
\multicolumn{2}{c|}{$\tau_t $} & 0.68 & $[ \, \tau_t - 5\%\tau_t, \, \tau_t + 5\%\tau_t \, ]$ & $1\%\tau_t$ \\ \midrule
\multicolumn{2}{c|}{$\tau_o $} & 0.92 & $[ \, \tau_o - 5\%\tau_o, \, \tau_o + 5\%\tau_o \, ]$ & $1\%\tau_o$ \\ \midrule
\multicolumn{2}{c|}{$\tau_l $} & 0.90 & $[ \, \tau_l - 5\%\tau_l, \, \tau_l + 5\%\tau_l \, ]$ &  $1\%\tau_l$ \\ 
\bottomrule[1.5pt]
\end{tabular}
}
\end{table}

In the evaluation, we measure the Precision, Recall, and F1-score, and compute the average variation of these metrics within the corresponding perturbation intervals. As shown in Table~\ref{Table: Sensitivity analysis for parameters.}, the average change across the interval remained minor, indicating that the value of parameters is stable and appropriate. Therefore, the sensitivity analysis verifies that the selected weights and thresholds provide robust performance under the perturbations.

\begin{table}[h]
\centering
\caption{Sensitivity analysis for parameters.}
\label{Table: Sensitivity analysis for parameters.}
\resizebox{0.48\textwidth}{!}{
\begin{tabular}{c|c|c|c}
\toprule[1.5pt]
\textbf{Parameters} & \textbf{$\overline{\Delta Precision}$} & \textbf{$\overline{\Delta Recall}$}  &  \textbf{$\overline{\Delta F1-Score}$}\\
\midrule[1.5pt]
$\boldsymbol{\omega}^{\,t}$ & 1.19\% & 0 & 0.68\%   \\
$\boldsymbol{\omega}^{\,o}$ & 0.70\%  & 1.05\% & 0.90\% \\ 
$\tau_t $ & 2.20\% & 0 & 1.35\% \\
$\tau_o $ & 3.35\% & 3.16\% & 2.96\% \\
$\tau_l $ & 0.83\% & 0.31\% &  0.57\% \\
\bottomrule[1.5pt]
\end{tabular}
}
\end{table}

\end{document}